\documentclass[12pt,thmsa]{article}
\usepackage{makeidx}
\usepackage{sw20lart}
\usepackage{babel}

\input{tcilatex}
\begin{document}

\author{S. Manoff \\
Bulgarian Academy of Sciences\\
Institute for Nuclear Research and Nuclear Energy\\
Department of Theoretical Physics\\
Blvd. Tzarigradsko Chaussee 72\\
1784 Sofia - Bulgaria}
\title{Relative velocity and relative acceleration induced by the torsion in $%
(L_n,g)$- and $U_n$-spaces }
\date{E-mail address: smanov@inrne.bas.bg }
\maketitle

\begin{abstract}
The influence of the torsion on the relative velocity and on the relative
acceleration between particles (points) in spaces with an affine connection
and a metric [$(L_n,g)$-spaces] and in (pseudo) Riemannian spaces with
torsion ($U_n$-spaces) is considered. Necessary and sufficient conditions as
well as only necessary and only sufficient conditions for vanishing
deformation, shear, rotation and expansion are found. The notion of relative
acceleration and the related to it notions of shear, rotation and expansion
accelerations induced by the torsion are determined. It is shown that the
kinematic characteristics induced by the torsion (shear acceleration,
rotation acceleration and expansion acceleration) could play the same role
as the kinematic characteristics induced by the curvature and can (under
given conditions) compensate their action as well as the action of external
forces. The change of the rate of change of the length of a deviation vector
field is given in explicit form for $(L_n,g)$- and $U_n$-spaces.

\textit{Short title}: Relative velocity and relative acceleration induced by
the torsion

\end{abstract}

\section{Introduction}

\subsection{Differential geometry and space-time geometry}

In the last years, the evolution of the relations between differential
geometry and space-time geometry has made some important steps toward
applications of more comprehensive differential-geometric structures in the
models of space-time than these used in (pseudo) Riemannian spaces without
torsion ($V_n$-spaces).

1. It has been proved recently that every differentiable manifold with one
affine connection and metrics [$(L_n,g)$-space] \cite{Hehl} could be used as
a model for a space-time. In it the equivalence principle (related to the
vanishing of the components of an affine connection at a point or on a curve
in a manifold) holds \cite{Iliev-1} $\div $ \cite{Iliev-7}, \cite{Hartley}.
Even if the manifold has two different (not only by sign) connections for
tangent and co-tangent vector fields [$(\overline{L}_n,g)$-space] \cite
{Manoff-0} \cite{Manoff-01} the principle of equivalence is fulfilled at
least for one of the two types of vector fields \cite{Manoff-1}. On this
grounds, every free moving spinless test particle in a suitable basic system
(frame of reference) \cite{Manoff-2} \cite{Manoff-3} will fulfil an equation
identical with the equation for a free moving spinless test particle in the
Newtons mechanics or in the special relativity. In $(\overline{L}_n,g)$- and 
$(L_n,g)$-spaces, this equation could be the autoparallel equation
[different from the geodesic equation in contrast to the case of (pseudo)
Riemannian spaces without torsion ($V_n$-spaces)].

2. There are evidences that $(L_n,g)$- and $(\overline{L}_n,g)$-spaces can
have similar structures as the $V_n$-spaces for describing dynamical systems
and the gravitational interaction. In such type of spaces one could use
Fermi-Walker transports \cite{Manoff-4} \cite{Manoff-5} \cite{Manoff-6}
conformal transports \cite{Manoff-7} \cite{Manoff-8} and different frames of
reference \cite{Manoff-3}. All these notions appear as generalizations of
the corresponding notions in $V_n$-spaces. For instance, in $(L_n,g)$- and $(%
\overline{L}_n,g)$-spaces a proper non-rotating accelerated observer's frame
of reference could be introduced by analogy of the same type of frame of
reference related to a Fermi-Walker transport \cite{Misner} in the Einstein
theory of gravitation (ETG).

3. All kinematic characteristics related to the notion of relative velocity 
\cite{Stephani} as well as the kinematic characteristics related to the
notion of relative acceleration have been worked out for $(L_n,g)$- and $(%
\overline{L}_n,g)$-spaces without changing their physical interpretation in $%
V_n$-spaces \cite{Manoff-9}. Necessary and sufficient conditions as well as
only necessary an only sufficient conditions for vanishing shear, rotation
and expansion acceleration induced by the curvature are found \cite{Manoff-9}%
. The last results are related to the possibility of building a theoretical
basis for construction of gravitational wave detectors on the grounds of
gravitational theories over $(L_n,g)$ and $(\overline{L}_n,g)$--spaces.

Usually, in the gravitational experiments the measurements of two basic
objects are considered \cite{Grishchuk}:

(a) The relative velocity between two particles (or points) related to the
rate of change of the length (distance) between them. The change of the
distance is supposed to be caused by the gravitational interaction.

(b) The relative accelerations between two test particles (or points) of a
continuous media. These accelerations are related to the curvature of the
space-time and supposed to be induced by a gravitational force.

Together with the accelerations induced by the curvature in $V_n$-spaces
accelerations induced by the torsion would appear in $U_n$-spaces, as well
as by torsion and by non-metricity in $(L_n,g)$- and $(\overline{L}_n,g)$%
-spaces.

In particular, in other models of a space-time [different from the (pseudo)
Riemannian spaces without torsion] the torsion has to be taken into account
if we consider the characteristics of the space-time.

4. On the one hand, until now, there are a few facts that the torsion could
induce some very small and unmesasurable effects in quantum mechanical
systems considered in spaces with torsion \cite{Laemmerzahl}. At the same
time, there are no evidences that the model's descriptions of interactions
on macro level should include the torsion as a necessary mathematical tool.
From this (physical) point of view the influence of the torsion on dynamical
systems could be ignored since it could not play an important role in the
description of physical systems in the theoretical gravitational physics. On
the other hand, from mathematical point of view (as we will try to show in
this paper), the role of the torsion in new theories for description of
dynamical systems could be important and not ignorable.

\subsection{Problems and results}

In this paper the influence of the torsion on the relative velocity and the
relative acceleration between particles moving in $(L_n,g)$- and $U_n$%
-spaces ($n=4$) is considered.

In Sec. 2. the notion of relative velocity as well as the related to it
notions of shear, rotation and expansion velocities, induced by the torsion,
are determined. The change of the length of a vector field (deviation vector
field) related to two moving particles is given in an explicit form for $%
(L_n,g)$- and $U_n$-spaces. It is shown that the existence of deformation,
shear, rotation and expansion (velocities), induced by the torsion, can
compensate (under given conditions) the action of these induced by external
forces. The necessary and sufficient conditions as well as only necessary
and only sufficient conditions for vanishing deformation, shear, rotation
and expansion are found. The vanishing of all these quantities could be
caused by the torsion.

In Sec. 3. the notion of relative acceleration and the related to it notions
of shear, rotation and expansion accelerations induced by the torsion are
found. It is shown that the existence of kinematic characteristics, induced
by the torsion, can compensate (under given conditions) the action of these
induced by the curvature or by external forces. The change of the rate of
change of the length of a deviation vector field is given in explicit form
for $(L_n,g)$- and $U_n$-spaces.

Concluding remarks comprise the final Sec. 4. The most considerations are
given in details (even in full details) for those readers who are not
familiar with the considered problems.

\section{Relative velocity and its kinematic characteristics induced by the
torsion}

Let us now recall some well known facts from differential geometry. Every
contravariant vector field $\xi \in T(M)$ over a differentiable manifold $M$
can be written by means of its projection along and orthogonal to a
contravariant non-null (nonisotropic) vector field $u$ in two parts - one
collinear to $u$ and one - orthogonal to $u$, i.e.

\begin{equation}  \label{Ch 8 2.7}
\xi =\frac le.u+h^u[g(\xi )]\text{ }=\frac le.u+\overline{g}[h_u(\xi )]\text{
,}
\end{equation}

\noindent where $\overline{g}[h_u(\xi )]:=\xi _{\perp }=g^{ik}\cdot
h_{kl}\cdot \xi ^l\cdot e_i\,$ is called deviation vector field and

\begin{equation}
\begin{array}{c}
l=g(\xi ,u)=g_{ij}\cdot \xi ^i\cdot u^j\text{ , } \\ 
\text{\thinspace \thinspace \thinspace \thinspace \thinspace \thinspace
\thinspace \thinspace \thinspace \thinspace \thinspace \thinspace \thinspace 
}h^u=\overline{g}-\frac 1e.u\otimes u\text{ ,\thinspace \thinspace
\thinspace }\xi =\xi ^i.\partial _i=\xi ^k.e_k\text{ , \thinspace \thinspace
\thinspace \thinspace \thinspace \thinspace \thinspace \thinspace \thinspace 
}h^u=h^{ij}.e_i.e_j\text{ ,} \\ 
\overline{g}(h_u)\overline{g}=h^u\text{ ,\thinspace \thinspace \thinspace
\thinspace \thinspace }h_u(\overline{g})(g)=h_u\text{ ,\thinspace \thinspace
\thinspace \thinspace \thinspace \thinspace }h^u(g)(\overline{g})=h^u\text{
,\thinspace \thinspace \thinspace \thinspace \thinspace \thinspace }%
g(h^u)g=h_u\text{ , } \\ 
h_u=g-\frac 1e\cdot g(u)\otimes g(u)\text{ ,} \\ 
\text{\thinspace \thinspace }\overline{g}=g^{ij}\cdot e_i.e_j\,\,\,\text{,
\thinspace \thinspace \thinspace }e_i.e_j=\frac 12(e_i\otimes e_j+e_j\otimes
e_i)\text{ , }g^{ij}=g^{ji}\text{ ,} \\ 
g=g_{ij}\cdot e^i.e^j\text{ , }e^i.e^j=\frac 12(e^i\otimes e^j+e^j\otimes
e^i)\text{ , \thinspace }g_{ij}=g_{ji}\text{ .}
\end{array}
\end{equation}

Therefore, the covariant derivative $\nabla _u\xi $ of the contravariant
vector field $\xi $ along a non-null vector field $u$ [as a result of the
action of the contravariant differential operator $\nabla _u:\xi \rightarrow
\nabla _u\xi \in T(M)$, $\nabla _u\xi =\xi ^i\,_{;j}\cdot u^j\cdot \partial
_i$, $\xi ^i{}_{;j}=\xi ^i{}_{,j}+\Gamma _{lj}^i\cdot \xi ^l$, $\Gamma
_{lj}^i$ are the components of the affine connection $\Gamma $ in a $(L_n,g)$%
-space] can be written in the form

\begin{equation}
\nabla _u\xi =\frac{\overline{l}}e\cdot u+\overline{g}[h_u(\nabla _u\xi )]%
\text{ ,\thinspace \thinspace \thinspace \thinspace \thinspace \thinspace
\thinspace \thinspace \thinspace \thinspace \thinspace \thinspace \thinspace
\thinspace \thinspace \thinspace \thinspace \thinspace }\overline{l}%
=g(\nabla _u\xi ,u)\text{ .}  \label{Ch 8 2.10}
\end{equation}

\subsection{Relative velocity in $(L_n,g)$-spaces}

The notion 
\index{relative velocity@relative velocity} \textit{relative velocity}
vector field (relative velocity) $_{rel}v$ can be defined in $(L_n,g)$%
-spaces (regardless of its physical interpretation) as the projection
[orthogonal to a non-null (nonisotropic) contravariant vector field $u$] of
the first covariant derivative $\nabla _u\xi $ (along the same non-null
vector field $u $) of (another) contravariant vector field $\xi $, i.e.

\begin{equation}
_{rel}v=%
\overline{g}(h_u(\nabla _u\xi ))=g^{ij}\cdot h_{jk}\cdot \xi ^k\text{ }%
_{;l}\cdot u^l\cdot e_i\,\,\text{,}\,\,\,\,\,\,\,\,\,e_i=\partial _i\text{
(in a co-ordinate basis),}  \label{Ch 8 2.1}
\end{equation}

\noindent where (the indices in a co-ordinate and in a non-co-ordinate basis
are written in both cases as Latin indices instead of Latin and Greek
indices)

\begin{equation}
\begin{array}{c}
h_u=g-\frac 1e.g(u)\otimes g(u)\text{ },\text{ }h_u=h_{ij}\cdot e^i.e^j\text{
, }\overline{g}=g^{ij}\cdot e_i.e_j, \\ 
\nabla _u\xi =\xi ^i\text{ }_{;j}\cdot u^j\cdot e_i\text{ , \thinspace
\thinspace \thinspace \thinspace \thinspace \thinspace \thinspace \thinspace
\thinspace \thinspace \thinspace \thinspace \thinspace \thinspace \thinspace
\thinspace }\xi ^i\text{ }_{;j}=e_j\xi ^i+\Gamma _{kj}^i\cdot \xi ^k\,\text{%
,\thinspace \thinspace \thinspace \thinspace \thinspace \thinspace
\thinspace \thinspace \thinspace \thinspace \thinspace \thinspace \thinspace
\thinspace \thinspace \thinspace \thinspace \thinspace \thinspace }%
\,\,\,\Gamma _{kj}^i\neq \Gamma _{jk}^i\text{ }, \\ 
e=g(u,u)=g_{ij}\cdot u^i\cdot u^j=u_i\cdot u^i\neq 0\text{ ,\thinspace
\thinspace \thinspace \thinspace \thinspace \thinspace \thinspace \thinspace
\thinspace \thinspace \thinspace \thinspace \thinspace \thinspace }%
g(u)=g_{ik}\cdot u^k\cdot e^i=u_i\cdot e^i\text{ ,} \\ 
\text{ }h_u(\nabla _u\xi )=h_{ij}\cdot \xi ^j\text{ }_{;k}\cdot u^k\cdot e^i%
\text{ ,\thinspace \thinspace \thinspace \thinspace \thinspace \thinspace
\thinspace \thinspace \thinspace \thinspace \thinspace \thinspace \thinspace
\thinspace \thinspace \thinspace \thinspace }h_{ij}=g_{ij}-\frac 1e\cdot
u_i\cdot u_j\text{ ,} \\ 
h_u(\nabla _u\xi )=h_{ij}\cdot \xi ^j\text{ }_{;k}\cdot u^k\cdot e^i\text{ .}
\end{array}
\label{Ch 8 2.2}
\end{equation}

In a co-ordinate basis $e_j\xi ^i=\xi ^i$ $_{,j}=\partial _j\xi ^i=\partial
\xi ^i/\partial x^j$, \thinspace \thinspace \thinspace $e^j=dx^j$, $%
\,\,\,\,\,\,e_i=\partial _i=\partial /\partial x^i$,$\,\,\,\,\,\,u=u^i\cdot
\partial _i$,\thinspace \textit{$e_{,k}=e_ke=\partial _ke$}.

Using the relation \cite{Yano} 
\index{Yano K.@Yano K.} between the Lie derivative $\pounds _\xi u$ and the
covariant derivative $\nabla _\xi u$

\begin{equation}  \label{Ch 8 2.11}
\pounds _\xi u=\nabla _\xi u-\nabla _u\xi -T(\xi ,u)%
\text{ ,\thinspace \thinspace \thinspace \thinspace \thinspace \thinspace
\thinspace \thinspace \thinspace \thinspace \thinspace \thinspace \thinspace
\thinspace \thinspace \thinspace }T(\xi ,u)=T_{ij}\,^k.\xi ^i.u^j.e_k\text{ ,%
}
\end{equation}

\noindent one can write $\nabla _u\xi $ in the form

\begin{equation}  \label{Ch 8 2.12}
\nabla _u\xi =(k)g(\xi )-\pounds _\xi u\text{ }=k[g(\xi )]-\pounds _\xi u%
\text{,}
\end{equation}

\noindent or, taking into account the above expression for $\xi $, in the
form $\nabla _u\xi =k[h_u(\xi )]+\frac le.a-\pounds _\xi u$, where

\begin{equation}
\begin{array}{c}
k[g(\xi )]=\nabla _\xi u-T(\xi ,u)\text{ , \thinspace \thinspace \thinspace }%
k=(u^i\text{ }_{;l}-T_{lk}\,^i.u^k).g^{lj}.e_i\otimes e_j=k^{ij}.e_i\otimes
e_j\text{ ,} \\ 
k[g(u)]=k(g)u=k^{ij}.g_{jk}.u^k.e_i=a=\nabla _uu=u^i\text{ }_{;j}.u^j.e_i%
\text{ .}
\end{array}
\label{Ch 8 2.13}
\end{equation}

$T_{ij}\,^k$ are the components of the torsion tensor $T$:

\[
\begin{array}{c}
T_{ij}\,^k=-T_{ji}\,^k=\Gamma _{ji}^k-\Gamma _{ij}^k-C_{ij}\text{ }^k\text{
\thinspace \thinspace \thinspace (in a non-co-ordinate basis }\{e_k\}\text{)
, } \\ 
\lbrack e_i,e_j]=\pounds _{e_i}e_j\text{ }=C_{ij}\text{ }^k.e_k\text{ ,} \\ 
\text{ \thinspace }T_{ij}\,^k=\Gamma _{ji}^k-\Gamma _{ij}^k\,\,\,\,\text{
(in a co-ordinate basis }\{\partial _k\}\text{ ) ,}
\end{array}
\]

For $h_u(\nabla _u\xi )$, it follows

\begin{equation}
h_u(\nabla _u\xi )=h_u(\frac le\cdot a-\pounds _\xi u)+h_u(k)h_u(\xi )\text{
,}  \label{Ch 8 2.15}
\end{equation}

\noindent where $h_u(k)h_u(\xi )=h_{ik}\cdot k^{kl}\cdot h_{lj}\cdot \xi
^j\cdot e^i$,\thinspace \thinspace \thinspace $h_u(u)=0$,$\,\,\,\,u(h_u)=0$, 
$h_u(k)h_u(u)=0$, $\,\,\,\,\,(u)h_u(k)h_u=0$.

If we introduce the abbreviation

\begin{equation}
d=h_u(k)h_u=h_{ik}\cdot k^{kl}\cdot h_{lj}\cdot e^i\otimes e^j=d_{ij}\cdot
e^i\otimes e^j\text{ ,}  \label{Ch 8 2.16}
\end{equation}

\noindent the expression for $_{rel}v$ can take the form

\[
_{rel}v=\overline{g}[h_u(\nabla _u\xi )]=\overline{g}(h_u)(\frac le\cdot
a-\pounds _\xi u)+\overline{g}[d(\xi )]= 
\]

\begin{equation}
=[g^{ik}\cdot h_{kl}\cdot (\frac le\cdot a^l-\pounds _\xi u^l)+g^{ik}\cdot
d_{kl}\cdot \xi ^l]\cdot e_i\text{ }=\,_{rel}v^i\cdot e_i\text{ ,}
\label{Ch 8 2.17}
\end{equation}

\noindent or

\begin{equation}
g(_{rel}v)=h_u(\nabla _u\xi )=h_u(\frac le\cdot a-\pounds _\xi u)+d(\xi )%
\text{ .}  \label{Ch 8 2.18}
\end{equation}

For the special case, when the vector field $\xi $ is orthogonal to $u$,
i.e. $\xi =\overline{g}[h_u(\xi )]$, and the Lie derivative of $u$ along $%
\xi $ is zero, i.e. $\pounds _\xi u=0$, then the relative velocity can be
written in the form $g(_{rel}v)=d(\xi )$ or in the form 
\[
_{rel}v=\overline{g}[d(\xi )]\text{.} 
\]

\textit{Remark}. All further calculations leading to a useful representation
of $d$ are quite straightforward. The problem here was the finding out a
representation of $h_u(\nabla _u\xi )$ in the form (\ref{Ch 8 2.15}) which
is not a trivial task.

\subsection{Deformation velocity, shear velocity, rotation velocity and
expansion velocity}

The covariant tensor field $d$ is a generalization for $(L_n,g)$-spaces of
the well known 
\index{deformation@deformation!deformation velocity@deformation velocity} 
\textit{deformation velocity }tensor for $V_n$-spaces \cite{Stephani} 
\index{Stephani H.@Stephani H.}, \cite{Kramer} 
\index{Kramer D.@Kramer D.}%
\index{Stephani H. (s. Kramer D.)@Stephani H. (s. Kramer D.)}%
\index{MacCallum M. (s. Kramer D.)@MacCallum M. (s. Kramer D.)}%
\index{Herlt E. (s. Kramer D.)@Herlt E. (s. Kramer D.)}. It is usually
represented by means of its three parts: the trace-free symmetric part,
called \textit{shear velocity }tensor (shear), the anti-symmetric part,
called 
\index{rotation@rotation!rotation velocity@rotation velocity} \textit{%
rotation velocity }tensor (rotation) and the trace part, in which the trace
is called 
\index{expansion@expansion!expansion velocity@expansion velocity} \textit{%
expansion velocity }(expansion)\textit{\ }invariant.

After some more complicated as for $V_n$-spaces calculations, the
deformation velocity tensor $d$ can be given in the form

\begin{equation}
d=h_u(k)h_u=h_u(k_s)h_u+h_u(k_a)h_u=\sigma +\omega +\frac 1{n-1}\cdot \theta
\cdot h_u%
\text{ ,}  \label{Ch 8 2.20}
\end{equation}

\noindent where $k_s=\,_sk^{ij}\cdot e_i.e_j$, $_sk^{ij}=\frac
12(k^{ij}+k^{ji})$, $_ak=\,_ak^{ij}\cdot e_i\wedge e_j$, $_ak^{ij}=\frac
12(k^{ij}-k^{ji})$, $e_i\wedge e_j=\frac 12(e_i\otimes e_j-e_j\otimes e_i)$

The tensor $\sigma $ is the 
\index{shear@shear!shear velocity@shear velocity} \textit{shear velocity}
tensor (shear) , 
\begin{equation}
\begin{array}{c}
\sigma =\,_sE-\,_sP=E-P-\frac 1{n-1}\cdot 
\overline{g}[E-P]\cdot h_u=\sigma _{ij}\cdot e^i.e^j= \\ 
=E-P-\frac 1{n-1}\cdot (\theta _o-\theta _1)\cdot h_u\text{ ,}
\end{array}
\label{Ch 8 2.21}
\end{equation}

\noindent where

\begin{equation}
\begin{array}{c}
_sE=E-\frac 1{n-1}\cdot \overline{g}[E]\cdot h_u\text{ , \thinspace
\thinspace \thinspace \thinspace \thinspace \thinspace \thinspace \thinspace
\thinspace \thinspace \thinspace \thinspace \thinspace \thinspace }\overline{%
g}[E]=g^{ij}\cdot E_{ij}=\theta _o\text{ ,} \\ 
E=h_u(\varepsilon )h_u\text{ , \thinspace \thinspace \thinspace \thinspace
\thinspace \thinspace \thinspace \thinspace \thinspace \thinspace \thinspace
\thinspace }k_s=\varepsilon -m\text{ , \thinspace \thinspace \thinspace
\thinspace \thinspace \thinspace \thinspace \thinspace \thinspace \thinspace
\thinspace }\varepsilon =\frac 12(u_{\text{ };l}^i\cdot g^{lj}+u_{\text{ }%
;l}^j\cdot g^{li})\cdot e_i.e_j\text{ ,} \\ 
m=\frac 12(T_{lk}\,^i\cdot u^k\cdot g^{lj}+T_{lk}\,^j\cdot u^k\cdot
g^{li})\cdot e_i.e_j\text{ .}
\end{array}
\label{Ch 8 2.22}
\end{equation}

The tensor $_sE$ is the \textit{torsion-free shear velocity} 
\index{shear@shear!torsion-free shear velocity@torsion-free shear velocity}%
\textit{\ }tensor, $_sP$ is the 
\index{shear@shear!shear velocity induced by the torsion@shear velocity
induced by the torsion} \textit{shear velocity} tensor \textit{induced by
the torsion},

\textit{
\begin{equation}
\begin{array}{c}
_sP=P-\frac 1{n-1}\cdot 
\overline{g}[P]\cdot h_u\text{ , \thinspace \thinspace \thinspace \thinspace
\thinspace \thinspace \thinspace \thinspace \thinspace \thinspace \thinspace
\thinspace \thinspace \thinspace \thinspace }\overline{g}[P]=g^{kl}\cdot
P_{kl}=\theta _1\text{, \thinspace \thinspace \thinspace \thinspace
\thinspace \thinspace }P=h_u(m)h_u\text{ ,} \\ 
\theta _1=T_{kl}\,^k\cdot u^l\text{ ,\thinspace \thinspace \thinspace
\thinspace \thinspace \thinspace \thinspace }\theta _o=u^n\text{ }%
_{;n}-\frac 1{2e}(e_{,k}\cdot u^k-g_{kl;m}\cdot u^m\cdot u^k\cdot u^l)\text{
,\thinspace \thinspace \thinspace }\theta =\theta _o-\theta _1\text{ . }
\end{array}
\label{Ch 8 2.25}
\end{equation}
}

The invariant $\theta $ is the 
\index{expansion@expansion!expansion velocity@expansion velocity} \textit{%
expansion velocity, $\theta _o$} is the 
\index{expansion@expansion!torsion-free expansion velocity@torsion-free
expansion velocity} \textit{torsion-free expansion velocity,} $\theta _1$ is
the \textit{expansion velocity induced by the torsion.}

The tensor $\omega $ is the 
\index{rotation@rotation!rotation velocity@rotation velocity}\textit{\
rotation velocity }tensor (rotation velocity),

\begin{equation}
\begin{array}{c}
\omega =h_u(k_a)h_u=h_u(s)h_u-h_u(q)h_u=S-Q%
\text{ ,} \\ 
s=\frac 12(u^k\text{ }_{;m}\cdot g^{ml}-u^l\text{ }_{;m}\cdot g^{mk})\cdot
e_k\wedge e_l\text{ ,} \\ 
q=\frac 12(T_{mn}\,^k\cdot g^{ml}-T_{mn}\,^l\cdot g^{mk})\cdot u^n\cdot
e_k\wedge e_l\text{ , \thinspace \thinspace }S=h_u(s)h_u\text{ , \thinspace
\thinspace }Q=h_u(q)h_u\text{ .}
\end{array}
\label{Ch 8 2.28}
\end{equation}

The tensor $S$ is the 
\index{rotation@rotation!torsion-free rotation velocity@torsion-free
rotation velocity} \textit{torsion-free rotation velocity} tensor, $Q$ is
the 
\index{rotation@rotation!rotation velocity induced by the torsion@rotation
velocity induced by the torsion}\textit{\ rotation velocity }tensor \textit{%
induced by the torsion.}

By means of the expressions for $\sigma $, $\omega $ and $\theta $ the
deformation velocity tensor $d$ can be decomposed in two parts: $d_0$ and $%
d_1$

\begin{equation}
d=d_o-d_1%
\text{ , \thinspace \thinspace \thinspace }d_o=\,_sE+S+\frac 1{n-1}.\theta
_o.h_u\text{ , \thinspace \thinspace \thinspace \thinspace }%
d_1=\,_sP+Q+\frac 1{n-1}.\theta _1.h_u\text{ ,}  \label{Ch 8 2.31}
\end{equation}

\noindent where $d_o$ is the 
\index{deformation@deformation!torsion-free deformation
velocity@torsion-free deformation velocity} \textit{torsion-free deformation
velocity} tensor and $d_1$ is the 
\index{deformation@deformation!deformation velocity induced by the
torsion@deformation velocity induced by the torsion} \textit{deformation
velocity }tensor \textit{induced by the torsion. }For the case of $V_n$%
-spaces $d_1=0$ ($_sP=0$, $Q=0$, $\theta _1=0$).

\textit{Remark}. The shear velocity tensor $\sigma $ and the expansion
velocity $\theta $ can also be written in the forms

\begin{equation}
\begin{array}{c}
\sigma =\frac 12\{h_u(\nabla _u%
\overline{g}-\pounds _u\overline{g})h_u-\frac 1{n-1}\cdot (h_u[\nabla _u%
\overline{g}-\pounds _u\overline{g}])\cdot h_u\}\text{ }= \\ 
=\frac 12\{h_{ik}\cdot (g^{kl}\text{ }_{;m}\cdot u^m-\pounds _ug^{kl})\cdot
h_{lj}-\frac 1{n-1}\cdot h_{kl}\cdot (g^{kl}\text{ }_{;m}\cdot u^m-\pounds
_ug^{kl})\cdot h_{ij}\}\cdot e^i.e^j\text{ ,}
\end{array}
\label{Ch 8 2.32}
\end{equation}
\[
\theta =\frac 12\cdot h_u[\nabla _u\overline{g}-\pounds _u\overline{g}%
]=\frac 12h_{ij}\cdot (g^{ij}\text{ }_{;k}\cdot u^k-\pounds _ug^{ij})\text{ .%
} 
\]

The physical interpretation of the velocity tensors $d$, $\sigma $, $\omega $%
, and of the invariant $\theta $ for the case of $V_4$-spaces \cite{Synge} 
\index{Synge J. L.@Synge J. L.}, \cite{Ehlers} 
\index{Ehlers J.@Ehlers J.}, can also be extended for $(L_4,g)$-spaces. In
this case the torsion plays an equivalent role in the velocity tensors as
the covariant derivative. It is easy to be seen that the existence of some
kinematic characteristics ($_sP$, $Q$, $\theta _1$) depends on the existence
of the torsion tensor field. They vanish if it is equal to zero (e.g. in $%
V_n $-spaces). On the other side, \textit{the kinematic characteristics,
induced by the torsion, can compensate the result of the action of the
torsion-free kinematic characteristics}. If $d=0$, $\sigma =0$, $\omega =0$, 
$\theta =0$, then we could have the relations $d_0=d_1$, $_sE=\,_sP$, $S=Q$, 
$\theta _0=\theta _1$ respectively leading to vanishing the relative
velocity $_{rel}v$ under the additional conditions $g(u,\xi )=l=0$ and $%
\pounds _\xi u=0$. Since $\nabla _u\xi =%
\frac{\overline{l}}e\cdot u+\,_{rel}v$ and $\overline{l}=g(\nabla _u\xi
,u)=ul-g(\xi ,a)-(\nabla _ug)(\xi ,u)=-g(\xi ,a)-(\nabla _ug)(\xi ,u)$ for $%
l=0$, $\nabla _u\xi =-(\nabla _ug)(\xi ,u)$ can be seen under the above
conditions as a measure for the non-metricity ($\nabla _ug$) of the
space-time if $a=0$ and $_{rel}v=0$.

\subsection{Special contravariant vector fields with vanishing kinematic
characteristics related to the relative velocity}

The explicit forms of the quantities $d$, $\sigma $, $\omega $, and $\theta $
related to the relative velocity can be used for finding conditions for
existence of special types of contravariant vector fields with vanishing
characteristics induced by the relative velocity.

\subsubsection{Contravariant vector fields with vanishing deformation
velocity $(d=0)$}

If we consider the explicit form for $d$%
\[
d:=h_u(k)h_u 
\]

\noindent we can prove the following propositions:

\textit{Proposition 1}. The necessary and sufficient condition for the
existence of a non-null contravariant vector field $u$ with vanishing
deformation velocity $(d=0)$ is the condition 
\[
k=\frac 1e\cdot \{a\otimes u+u\otimes [g(u)](k)\}-\frac 1{2e^2}\cdot
[ue-(\nabla _ug)(u,u)]\cdot u\otimes u\text{ ,} 
\]

\noindent or in a co-ordinate basis 
\begin{eqnarray*}
k^{ij} &=&\frac 1e\cdot (a^i\cdot u^j+u^i\cdot k^{lj}\cdot g_{lm}\cdot u^m)
\\
&&-\frac 1{2e^2}\cdot (e_{,k}\cdot u^k-g_{kl;m}\cdot u^m\cdot u^k\cdot
u^l)\cdot u^i\cdot u^j\text{ .}
\end{eqnarray*}

Proof. 1. Necessity. From $d=h_u(k)h_u$, after writing the explicit form of $%
h_u$, it follows that 
\begin{eqnarray*}
d &=&g(k)g-\frac 1e\cdot g(u)\otimes [g(u)](k)g-\frac 1e\cdot
g(k)[g(u)]\otimes g(u) \\
&&+\frac 1{e^2}\cdot [g(u)](k)[g(u)]\cdot u\otimes u\text{ .}
\end{eqnarray*}

Since $(k)[g(u)]=a=\nabla _uu$, it follows further that 
\[
\lbrack g(u)](k)[g(u)]=[g(u)](a)=g(u,a)=\frac 12\cdot [ue-(\nabla _ug)(u,u)]%
\text{ .} 
\]

Therefore, 
\begin{eqnarray*}
d &=&0:g(k)g=\frac 1e\cdot \{g(u)\otimes [g(u)](k)g+g(a)\otimes g(u)\} \\
&&-\frac 1{e^2}\cdot g(u,a)\cdot g(u)\otimes g(u)\text{ .}
\end{eqnarray*}

From $\overline{g}(g(k)g)\overline{g}=k$, we obtain 
\begin{eqnarray*}
g(k)g &=&\frac 1e\cdot \{a\otimes u+u\otimes [g(u)](k)\} \\
&&-\frac 1{2\cdot e^2}\cdot [ue-(\nabla _ug)(u,u)]\cdot u\otimes u\text{ .}
\end{eqnarray*}

2. Sufficiency. From the explicit form of $k$, it follows that 
\begin{eqnarray*}
g(k)g &=&\frac 1e\cdot g(u)\otimes [g(u)](k)g+\frac 1e\cdot
g(k)[g(u)]\otimes g(u) \\
&&-\frac 1{e^2}\cdot g(u,a)\cdot g(u)\otimes g(u)\text{ }
\end{eqnarray*}

\noindent which is identical to $h_u(k)h_u=d=0$.

\textit{Special case:} $\nabla _uu=a:=0$, $\nabla _\xi g:=0$ for $\forall
\xi \in T(M)$ ($U_n$-space), $ue=0:e=\,$const.$\,\neq 0$ ($u$ is normalized,
non-null contravariant vector field). 
\[
d=0:k=\frac 1e\cdot u\otimes [g(u)](k)\text{ ,} 
\]
\[
(k)[g(\xi )]=\frac 1e\cdot u\otimes [g(u)](k)[g(\xi )]=\frac 1e\cdot
[g(u)](k)[g(\xi )]\cdot u\text{ ,} 
\]
\begin{eqnarray*}
(k)[g(\xi )] &=&(u^i\,_{;l}-T_{lk}\,^i\cdot u^k)\cdot g^{lm}\cdot
g_{mj}\cdot \xi ^j\cdot \partial _i=\nabla _\xi u-T(\xi ,u)= \\
&=&\nabla _u\xi -\pounds _\xi u\text{ ,}
\end{eqnarray*}
\begin{equation}
_{rel}v=\overline{g}[h_u(\nabla _u\xi )]=-\,\overline{g}(h_u)(\pounds _\xi u)%
\text{ for }\forall \xi \in T(M)\text{ \thinspace \thinspace .}
\label{Ch 8 2.33a}
\end{equation}

\textit{Proposition 2. }A sufficient condition for the existence of a
non-null contravariant vector field with vanishing deformation velocity ($%
d=0 $) is the condition 
\[
k=0\text{ ,} 
\]

\noindent equivalent to the condition 
\[
\nabla _\xi u=T(\xi ,u)\text{ \thinspace \thinspace \thinspace for }\forall
\xi \in T(M)\text{ ,} 
\]

\noindent or in a co-ordinate basis 
\[
k^{ij}=0:u^i\,_{;j}=T_{jk}\,^i\cdot u^k\text{ .} 
\]

Proof. From $k=0$ and $(k)[g(\xi )]=\nabla _\xi u-T(\xi ,u)$ for $\forall
\xi \in T(M)$, it follows that $\nabla _\xi u-T(\xi ,u)=0$ or in a
co-ordinate basis $u^i\,_{;j}-T_{jk}\,^i\cdot u^k=0$. In this case $\pounds
_\xi u=\nabla _\xi u-\nabla _u\xi -T(\xi ,u)=-\nabla _u\xi $.

\textit{Corollary}. A deformation free contravariant vector field $u$ with $%
k=0$ is an auto-parallel contravariant vector field.

Proof. It follows immediately from the condition $\nabla _\xi u=T(\xi ,u)$
and for $\xi =u$ that $\nabla _uu=a=0$.

\textit{Proposition 3}. The necessary condition for the existence of a
deformation free contravariant vector field $u$ with $k=0$ is the condition 
\[
\lbrack R(u,v)]\xi =[\pounds \Gamma (u,v)]\xi \text{ \thinspace \thinspace
\thinspace \thinspace \thinspace \thinspace \thinspace \thinspace \thinspace
for\thinspace \thinspace \thinspace \thinspace }\forall \xi ,v\in T(M)\text{
,} 
\]

\noindent or in a co-ordinate basis 
\[
R^k\,_{ilj}\cdot u^l=\pounds _u\Gamma _{ij}^k\text{ .} 
\]

Proof. By the use of the explicit form of the curvature operator $R(u,v)$
acting on a contravariant vector field $\xi $%
\[
\lbrack R(u,v)]\xi =\nabla _u\nabla _v\xi -\nabla _v\nabla _u\xi -\nabla
_{\pounds _uv}\xi \text{ , \thinspace \thinspace \thinspace \thinspace
\thinspace \thinspace \thinspace \thinspace \thinspace \thinspace \thinspace
\thinspace \thinspace }\xi \text{, }v\text{, }u\in T(M)\text{ ,} 
\]

\noindent and the explicit form of the deviation operator $\pounds \Gamma
(u,v)$ acting on a contravariant vector field $\xi $%
\[
\lbrack \pounds \Gamma (u,v)]\xi =\pounds _u\nabla _v\xi -\nabla _v\pounds
_u\xi -\nabla _{\pounds _uv}\xi 
\]

\noindent we obtain under the condition $\nabla _\xi u=T(\xi ,u)$
(equivalent to the condition $\pounds _u\xi =\nabla _u\xi $) 
\begin{eqnarray*}
\lbrack \pounds \Gamma (u,v)]\xi &=&\pounds _u\nabla _v\xi -\nabla _v\pounds
_u\xi -\nabla _{\pounds _uv}\xi = \\
&=&\nabla _u\nabla _v\xi -\nabla _{\nabla _v\xi }u-T(u,\nabla _v\xi )-\nabla
_v\nabla _u\xi -\nabla _{\pounds _uv}\xi = \\
&=&\nabla _u\nabla _v\xi -\nabla _v\nabla _u\xi -\nabla _{\pounds _uv}\xi
-[\nabla _{\nabla _v\xi }u+T(u,\nabla _v\xi )]= \\
&=&[R(u,v)]\xi -[\nabla _{\nabla _v\xi }u+T(u,\nabla _v\xi )]\text{ .}
\end{eqnarray*}

Since 
\[
\nabla _{\nabla _v\xi }u+T(u,\nabla _v\xi )=0\text{ \thinspace \thinspace
\thinspace \thinspace \thinspace \thinspace \thinspace for\thinspace
\thinspace \thinspace \thinspace \thinspace \thinspace \thinspace \thinspace 
}\forall v,\xi \in T(M)\text{ , } 
\]

\noindent we have 
\[
\lbrack R(u,v)]\xi =[\pounds \Gamma (u,v)]\xi \text{ \thinspace \thinspace
\thinspace \thinspace \thinspace \thinspace for\thinspace \thinspace
\thinspace \thinspace \thinspace \thinspace }\forall v,\xi \in T(M)\text{
\thinspace \thinspace . } 
\]

The last condition appears as the integrability condition for the equation
for $u$%
\[
\nabla _\xi u=T(\xi ,u)\text{ \thinspace \thinspace \thinspace \thinspace
for \thinspace \thinspace \thinspace \thinspace }\forall \xi \in T(M)\text{
\thinspace \thinspace .} 
\]

\textit{Proposition 4.} A deformation-free contravariant non-null vector
field $u$ with $k=0$ is an auto-parallel non-null shear-free ($\sigma =0$),
rotation-free ($\omega =0$) and expansion-free ($\theta =0$) contravariant
vector field with vanishing deformation acceleration ($A=0$) \cite{Manoff-9}.

Proof. If $k=k^{ij}\cdot \partial _i\otimes \partial _j=0$ and $\nabla
_uu=a=0$, then $k_s=k^{(ij)}\cdot \partial _i.\partial _j=\frac 12\cdot
(k^{ij}+k^{ji})\cdot \partial _i.\partial _j=0$, and $k_a=k^{[ij]}\cdot
\partial _i\wedge \partial _j=\frac 12\cdot (k^{ij}-k^{ji})\cdot \partial
_i\wedge \partial _j=0$. Therefore, $\sigma =h_u(k_s)h_u-\frac 1{n-1}\cdot 
\overline{g}[h_u(k_s)h_u]\cdot h_u=0$, $\theta =\overline{g}[h_u(k_s)h_u]=0$%
, and $\omega =h_u(k_a)h_u=0$. From the explicit form of the deformation
acceleration $A$, it follows that $A=0$.

From the identity for the Riemannian tensor $R^i\,_{jkl}$%
\begin{eqnarray*}
R^i\,_{jkl}+R^i\,_{ljk}+R^i\,_{klj} &\equiv
&T_{jk}\,^i\,_{;l}+T_{lj}\,^i\,_{;k}+T_{kl}\,^i\,_{;j}+ \\
&&+T_{jk}\,^m\cdot T_{ml}\,^i+T_{lj}\,^m\cdot T_{mk}\,^i+T_{kl}\,^m\cdot
T_{mj}\,^i\text{ ,}
\end{eqnarray*}

\noindent after contraction with $g_i^l$ and summation over $l$ we obtain 
\begin{eqnarray*}
R_{jk}-R_{kj}+R^i\,_{ijk} &\equiv
&T_{jk}\,^i\,_{;i}+T_{ij}\,^i\,_{;k}-T_{ik}\,^i\,_{;j}+ \\
&&+T_{jk}\,^m\cdot T_{mi}\,^i+T_{ij}\,^m\cdot T_{mk}\,^i-T_{ik}\,^m\cdot
T_{mj}\,^i\text{ .}
\end{eqnarray*}

If we introduce the abbreviations 
\[
_aR_{jk}:=\frac 12\cdot (R_{jk}-R_{kj})\text{ , \thinspace \thinspace
\thinspace \thinspace \thinspace \thinspace \thinspace \thinspace \thinspace
\thinspace \thinspace \thinspace }T_{ji}\,^i:=T_j\text{ ,\thinspace
\thinspace \thinspace \thinspace \thinspace \thinspace \thinspace \thinspace
\thinspace \thinspace \thinspace \thinspace } 
\]

\noindent where \thinspace $T_{ik}\,^i=-T_{ki}\,^i=-T_k$, then the last
expression for $R_{jk}$ can be written in the form 
\begin{eqnarray*}
2\cdot \,_aR_{jk} &\equiv
&-R^i\,_{ijk}+T_{jk}\,^i\,_{;i}+T_{ij}\,^i\,_{;k}-T_{ik}\,^i\,_{;j}+ \\
&&+T_{jk}\,^m\cdot T_m+T_{ij}\,^m\cdot T_{mk}\,^i-T_{ik}\,^m\cdot T_{mj}\,^i%
\text{ .}
\end{eqnarray*}

Therefore, $_aR_{ij}\cdot u^j$ can be written in the form 
\begin{eqnarray*}
2\cdot \,_aR_{ij}\cdot u^j+R^i\,_{ijk}\cdot u^j &=&T_{k;j}\cdot
u^j-T_{j;k}\cdot u^j+T_{jk}\,^i\,_{;i}\cdot u^j+ \\
&&+T_m\cdot T_{jk}\,^m\cdot u^j+T_{ij}\,^m\cdot u^j\cdot
T_{mk}\,^i-T_{ik}\,^m\cdot T_{mj}\,^i\cdot u^j\text{ .}
\end{eqnarray*}

From the other side, from $u^i\,_{;j}=T_{jl}\,^i\cdot u^l$ and $%
a^i=u^i\,_{;j}\cdot u^j=0$, we have 
\[
u^i\,_{;j;k}=T_{jl}\,^i\,_{;k}\cdot u^l+T_{jm}\,^i\cdot T_{kl}\,^m\cdot u^l%
\text{ ,} 
\]
\begin{eqnarray*}
u^i\,_{;j;k}-u^i\,_{;k;j} &=&-u^l\cdot R^i\,_{ljk}+T_{jk}\,^m\cdot
T_{ml}\,^i\cdot u^l= \\
&=&T_{jl}\,^i\,_{;k}\cdot u^l+T_{jm}\,^i\cdot T_{kl}\,^m\cdot u^l- \\
&&-T_{kl}\,^i\,_{;j}\cdot u^l-T_{km}\,^i\cdot T_{jl}\,^m\cdot u^l\,\text{ ,}
\end{eqnarray*}
\begin{eqnarray*}
u^l\cdot R^i\,_{ljk} &=&T_{jk}\,^m\cdot T_{ml}\,^i\cdot
u^l+T_{kl}\,^i\,_{;j}\cdot u^l-T_{jl}\,^i\,_{;k}\cdot u^l+ \\
&&+T_{km}\,^i\cdot T_{jl}\,^m\cdot u^l-T_{jm}\,^i\cdot T_{kl}\,^m\cdot u^l%
\text{ \thinspace \thinspace ,}
\end{eqnarray*}
\[
R_{lj}\cdot u^l=-T_{l;j}\cdot u^l-T_{jl}\,^i\,_{;i}\cdot u^l-T_m\cdot
T_{jl}\,^m\cdot u^l\text{ \thinspace \thinspace \thinspace ,} 
\]
\[
R_{lj}\cdot u^l\cdot u^j=\,_sR_{lj}\cdot u^l\cdot u^j=I=-T_{l;j}\cdot
u^l\cdot u^j=-(T_i\cdot u^i)_{;j}\cdot u^j=\dot{\theta}_1\text{ . } 
\]

By the use of the decompositions $R_{ij}=\,_aR_{ij}+\,_sR_{ij}$, $%
R_{ij}\cdot u^j=\,_aR_{ij}\cdot u^j+\,_sR_{ij}\cdot u^j$, and the above
expression for $R_{lj}\cdot u^l$, we can find the following relations 
\[
2\cdot \,_aR_{jk}\cdot u^j=T_{k;j}\cdot u^j-T_{j;k}\cdot u^j-2\cdot
T_{kj}\,^i\,_{;i}\cdot u^j-2\cdot T_m\cdot T_{kj}\,^m\cdot u^j\text{ ,} 
\]
\[
R^i\,_{ijk}\cdot u^j=(T_{kj}\,^i\,_{;i}+T_m\cdot T_{kj}\,^m+T_{ij}\,^m\cdot
T_{mk}\,^i-T_{ik}\,^m\cdot T_{mj}\,^i)\cdot u^j\text{ ,} 
\]
\[
2\cdot \,_sR_{jk}\cdot u^j=-(T_{j;k}+T_{k;j})\cdot u^j\text{ }. 
\]

It follows that in a $(\overline{L}_n,g)$-space the projections of the
symmetric part of the Ricci tensor on the non-null contravariant vector
field $u$ with $k=0$ is depending on the covariant derivatives of $T_i$
(respectively on the torsion $T_{ik}\,^l$) and not on the torsion $%
T_{ik}\,^l $ itself.

\subsubsection{Contravariant non-null (nonisotropic) vector fields with
vanishing shear velocity $(\sigma =0)$}

If we consider the explicit form of the shear velocity tensor (shear
velocity, shear) 
\[
\sigma =h_u(k_s)h_u-\frac 1{n-1}\cdot \overline{g}[h_u(k_s)h_u]\cdot h_u 
\]

\noindent we can prove the following propositions:

\textit{Proposition 5.} The necessary and sufficient condition for the
existence of a shear-free non-null contravariant vector field is the
condition 
\begin{eqnarray*}
k_s &=&\frac 1{2\cdot e}\cdot \{u\otimes a+a\otimes u+u\otimes
[g(u)](k)+[g(u)](k)\otimes u- \\
&&-\frac 1e\cdot [ue-(\nabla _ug)(u,u)]\cdot u\otimes u\}+ \\
&&+\frac 1{n-1}\cdot \theta \cdot h^u\text{,}
\end{eqnarray*}

\noindent or in a co-ordinate basis 
\begin{eqnarray*}
h_s^{ij} &=&\frac 1{2\cdot e}\cdot \{u^i\cdot a^j+u^j\cdot a^i+u^i\cdot g_{%
\overline{m}\overline{n}}\cdot u^n\cdot k^{mj}+u^j\cdot g_{\overline{m}%
\overline{n}}\cdot u^n\cdot k^{mi}- \\
&&-\frac 1e\cdot [e_{,k}\cdot u^k-g_{km;n}\cdot u^n\cdot u^{\overline{k}%
}\cdot u^{\overline{m}}]\cdot u^i\cdot u^j\}+\frac 1{n-1}\cdot \theta \cdot
h^{ij}\text{ .}
\end{eqnarray*}

Proof. 1. Necessity. From $\sigma =0$, it follows that $h_u(k_s)h_u=\frac
1{n-1}\cdot \overline{g}[h_u(k_s)h_u]\cdot h_u=\frac 1{n-1}\cdot \theta
\cdot h_u$. Further, from the explicit form of $h_u$ and $k_s$, it follows
that 
\begin{eqnarray*}
h_u(k_s)h_u &=&\frac 1{n-1}\cdot \theta \cdot h_u=g(k_s)g-\frac 1e\cdot
\{g(u)\otimes [g(u)](k_s)g+g(k_s)[g(u)]\otimes g(u)\}+ \\
&&+\frac 1{e^2}\cdot [g(u)](k_s)[g(u)]\cdot g(u)\otimes g(u)\text{ , }
\end{eqnarray*}

\noindent or 
\begin{eqnarray*}
k_s &=&\frac 1e\cdot \{u\otimes [g(u)](k_s)+(k_s)[g(u)]\otimes u\}-\frac
1{e^2}\cdot [g(u)](k_s)[g(u)]\cdot u\otimes u+ \\
&&+\frac 1{n-1}\cdot \theta \cdot \overline{g}(h_u)\overline{g}\text{ .}
\end{eqnarray*}

Since $[g(u)](k_s)=(k_s)[g(u)]$, $(k_s)[g(u)]=\frac 12\cdot
\{(k)[g(u)]+[g(u)](k)\}$, $(k)[g(u)]=a$, $(k_s)[g(u)]=\frac 12\cdot
\{a+[g(u)](k)\}$, $[g(u)](k_s)[g(u)]=[g(u)](k)[g(u)]=g(u,a)=\frac 12\cdot
[ue-(\nabla _ug)(u,u)]$, $\theta =\overline{g}[h_u(k_s)h_u]=\overline{g}[%
h_u(k)h_u]$, and $\overline{g}(h_u)\overline{g}=h^u$, the explicit form of $%
k_s$ can be found as 
\begin{eqnarray*}
k_s &=&\frac 1{2\cdot e}\cdot \{u\otimes a+a\otimes u+u\otimes
[g(u)](k)+[g(u)](k)\otimes u- \\
&&-\frac 1e\cdot [ue-(\nabla _ug)(u,u)]\cdot u\otimes u\}+ \\
&&+\frac 1{n-1}\cdot \theta \cdot h^u\text{ \thinspace \thinspace \thinspace
.}
\end{eqnarray*}

2. Sufficiency. From the last expression and the above relations, it follows
that $h_u(k_s)h_u=\frac 1{n-1}\cdot \theta \cdot h_u$, and therefore $\sigma
=0$.

\textit{Proposition 6.} A sufficient condition for the existence of a
non-null vector field with vanishing shear velocity ($\sigma =0$) and
expansion velocity ($\theta =0$) is the condition 
\[
h_u(k_s)h_u=0\text{ ,} 
\]

\noindent identical with the condition 
\begin{eqnarray*}
k_s &=&\frac 1{2\cdot e}\cdot \{u\otimes a+a\otimes u+u\otimes
[g(u)](k)+[g(u)](k)\otimes u- \\
&&-\frac 1e\cdot [ue-(\nabla _ug)(u,u)]\cdot u\otimes u\}
\end{eqnarray*}

Proof. If $h_u(k_s)h_u=0$, then $\theta =\overline{g}[h_u(k_s)h_u]=0$.
Therefore, $\sigma =h_u(k_s)h_u-\frac 1{n-1}\cdot \theta \cdot h_u=0$.

\textit{Corollary}. If $h_u(k_s)h_u=0$, then 
\[
g[k_s]=\frac 1{2\cdot e}\cdot [ue-(\nabla _ug)(u,u)]\text{ .} 
\]

Proof. It follows from the above proposition that 
\begin{eqnarray*}
\theta &=&g[k_s]-\frac 1e\cdot g(u,a)=0\text{ ,} \\
g[k_s] &=&\frac 1e\cdot g(u,a)=\frac 1{2\cdot e}\cdot [ue-(\nabla _ug)(u,u)]%
\text{ .}
\end{eqnarray*}

\subsubsection{Contravariant non-null vector fields with vanishing rotation
velocity $(\omega =0)$}

If we consider the explicit form for $\omega $%
\[
\omega =h_u(k_a)h_u 
\]

\noindent we can prove the following propositions:

\textit{Proposition 7.} The necessary and sufficient condition for the
existence of a contravariant non-null vector field with vanishing rotation
velocity $(\omega =0)$ is the condition 
\begin{equation}
k_a=\frac 1e\cdot \{u\otimes [g(u)](k_a)-[g(u)](k_a)\otimes u\}\text{ ,}
\label{10.1a}
\end{equation}

\noindent or in a co-ordinate basis 
\begin{equation}
k_a^{ij}=\frac 1e\cdot g_{\overline{m}\overline{n}}\cdot u^n\cdot (u^i\cdot
k_a^{mj}-u^j\cdot k_a^{mi})\text{ \thinspace \thinspace \thinspace
\thinspace .}  \label{10.1}
\end{equation}

Proof. 1. Necessity. Form $h_u(k_a)h_u=0$, it follows that 
\begin{eqnarray*}
h_u(k_a)h_u &=&0=g(k_a)g-\frac 1e\cdot g(u)\otimes [g(u)](k_a)g-\frac
1e\cdot g(k_a)[g(u)]\otimes g(u)+ \\
&&+\frac 1{e^2}\cdot [g(u)](k_a)[g(u)]\cdot g(u)\cdot g(u)\text{ .}
\end{eqnarray*}

Since 
\[
\lbrack g(u)](k_a)[g(u)]=g_{\overline{i}\overline{m}}\cdot u^m\cdot
k_a^{ij}\cdot g_{\overline{j}\overline{n}}\cdot u^n=-g_{\overline{i}%
\overline{m}}\cdot u^m\cdot k_a^{ij}\cdot g_{\overline{j}\overline{n}}\cdot
u^n\text{ ,} 
\]

\noindent we have $[g(u)](k_a)[g(u)]=0$. Therefore, 
\[
g(k_a)g=\frac 1e\cdot \{g(u)\otimes [g(u)](k_a)g+g(k_a)[g(u)]\otimes g(u)\}%
\text{ .} 
\]

From the last expression and from the relation $\overline{g}[g(k_a)g]%
\overline{g}=k_a$, it follows that 
\begin{eqnarray*}
k_a &=&\frac 1e\cdot \{u\otimes [g(u)](k_a)+(k_a)[g(u)]\otimes u\}= \\
&=&\frac 1e\cdot \{u\otimes [g(u)](k_a)-[g(u)](k_a)\otimes u\}\text{ ,}
\end{eqnarray*}

\noindent because of $(k_a)[g(u)]=-$ $[g(u)](k_a)$. In a co-ordinate basis
we obtain (\ref{10.1}).

2. Sufficiency. From (\cite{10.1a}) we have 
\[
g(k_a)g=\frac 1e\cdot \{g(u)\otimes [g(u)](k_a)g+g(k_a)[g(u)]\otimes g(u)\}%
\text{ ,} 
\]

\noindent which is identical to $h_u(k_a)h_u=0$.

On the other hand, after direct computations, it follows that 
\[
(k_a)[g(u)]=\frac 12\cdot \{(k)[g(u)]-[g(u)](k)\}\text{ .} 
\]

Since $(k)[g(u)]=a$, we have the relation 
\[
(k_a)[g(u)]=\frac 12\cdot \{a-[g(u)](k)\}\text{ .} 
\]

Then 
\[
k_a=\frac 1{2\cdot e}\cdot \{a\otimes u-u\otimes a+u\otimes
[g(u)](k)-[g(u)](k)\otimes u\}\text{ .} 
\]

\textit{Proposition 8.} A sufficient condition for the existence of a
contravariant non-null vector field with vanishing rotation velocity $%
(\theta =0)$ is the condition 
\[
k_a=0\text{ .} 
\]

Proof. If $k_a=0$, then it follows directly from $\omega =h_u(k_a)h_u$ that $%
\omega =0$.

In a co-ordinate basis $k_a$ is equivalent to the expression 
\[
u^i\,_{;l}\cdot g^{lj}-u^j\,_{;l}\cdot g^{li}=(T_{lm}\,^i\cdot
g^{lj}-T_{lm}\,^j\cdot g^{li})\cdot u^m\text{ .} 
\]

On the other side, after multiplying the last expression with $g_{\overline{j%
}\overline{k}}\cdot u^k$ and summarizing over $j$, we obtain 
\[
a^i=u^i\,_{;k}\cdot u^k=g_{\overline{j}\overline{k}}\cdot u^k\cdot
(u^j\,_{;l}-T_{lm}\,^j\cdot u^m)\cdot g^{li}=g_{\overline{j}\overline{k}%
}\cdot u^k\cdot k^{ji}\text{ ,} 
\]

\noindent or in a form 
\[
a=[g(u)](k)\text{ .} 
\]

\textit{Proposition 9}. The necessary condition for $k_a=0$ is the condition 
\[
a=[g(u)](k)\,\,\,\text{.} 
\]

Proof. From $k_a=0$ and $(k_a)[g(u)]=\frac 12\cdot \{a-[g(u)](k)\}$, it
follows that $a=[g(u)](k)$.

If the rotation velocity $\omega $ vanishes $(\omega =0)$ for an
auto-parallel $(\nabla _uu=0)$ contravariant non-null vector field $u$, then
the rotation acceleration tensor $W$ will have the form \cite{Manoff-9} 
\[
W=\frac 12\cdot [h_u(\nabla _u\overline{g})\sigma -\sigma (\nabla _u%
\overline{g})h_u]\text{ .} 
\]

From the last expression it is obvious that the nonmetricity $(\nabla
_ug\neq 0)$ in a $(\overline{L}_n,g)$-space is responsible for nonvanishing
the rotation acceleration $W$.

\subsection{Relative velocity and change of the length of a contravariant
vector field over a $(L_n,g)$-space}

Let us now consider the influence of the kinematic characteristics related
to the relative velocity and respectively to the relative velocity, induced
by the torsion, upon the change of the length of a contravariant vector
field.

Let $l_\xi =\,\mid g(\xi ,\xi )\mid ^{\frac 12}$ be the length of a
contravariant vector field $\xi $. The rate of change $ul_\xi $ of $l_\xi $
along a contravariant vector field $u$ can be expressed in the form $\pm
\,2.l_\xi .(ul_\xi )=(\nabla _ug)(\xi ,\xi )+2g(\nabla _u\xi ,\xi )$. By the
use of the projections of $\xi $ and $\nabla _u\xi $ along and orthogonal to 
$u$ we can find the relations 
\[
\begin{array}{c}
2g(\nabla _u\xi ,\xi )=2.\frac le.g(\nabla _u\xi ,u)+2g(_{rel}v,\xi _{\perp
}) \text{ ,} \\ 
(\nabla _ug)(\xi ,\xi )=(\nabla _ug)(\xi _{\perp },\xi _{\perp })+2.\frac
le.(\nabla _ug)(\xi _{\perp },u)+\frac{l^2}{e^2}.(\nabla _ug)(u,u)\text{ .}
\end{array}
\]

Then, it follows for $\pm \,2.l_\xi .(ul_\xi )$ the expression 
\begin{equation}  \label{Ch 8 2.34}
\begin{array}{c}
\pm \,2.l_\xi .(ul_\xi )=(\nabla _ug)(\xi _{\perp },\xi _{\perp })+2.\frac
le.[(\nabla _ug)(\xi _{\perp },u)+g(\nabla _u\xi ,u)]+ \\ 
+\,\frac{l^2}{e^2}.(\nabla _ug)(u,u)+2g(_{rel}v,\xi _{\perp })\text{ ,}
\end{array}
\end{equation}

where 
\begin{equation}  \label{Ch 8 2.35}
g(_{rel}v,\xi _{\perp })=\frac le.h_u(a,\xi _{\perp })+h_u(\pounds _u\xi
,\xi _{\perp })+d(\xi _{\perp },\xi _{\perp })\text{ ,}
\end{equation}
\begin{equation}  \label{Ch 8 2.36}
d(\xi _{\perp },\xi _{\perp })=\sigma (\xi _{\perp },\xi _{\perp })+\frac
1{n-1}.\theta .l_{\xi _{\perp }}^2\text{ .}
\end{equation}

For finding out the last two expressions the following relations have been
used: 
\begin{equation}  \label{Ch 8 2.37}
g(\overline{g}(h_u)a,\xi _{\perp })=h_u(a,\xi _{\perp })\text{ ,\thinspace
\thinspace \thinspace \thinspace \thinspace \thinspace \thinspace }g(%
\overline{g}(h_u)(\pounds _u\xi ),\xi _{\perp })=h_u(\pounds _u\xi ,\xi
_{\perp })\text{ ,}
\end{equation}
\begin{equation}  \label{Ch 8 2.38}
g(\overline{g}[d(\xi )],\xi _{\perp })=d(\xi _{\perp },\xi _{\perp })\text{
,\thinspace \thinspace \thinspace \thinspace \thinspace \thinspace
\thinspace \thinspace }d(\xi )=d(\xi _{\perp })\text{ .}
\end{equation}

\textit{Special case}: $g(u,\xi )=l:=0:\xi =\xi _{\perp }$. 
\begin{equation}  \label{Ch 8 2.39}
\pm \,2.l_{\xi _{\perp }}.(ul_{\xi _{\perp }})=(\nabla _ug)(\xi _{\perp
},\xi _{\perp })+2g(_{rel}v,\xi _{\perp })\text{ .}
\end{equation}

\textit{Special case}: $V_n$-spaces: $\nabla _\eta g=0$ for $\forall \eta
\in T(M)$ ($g_{ij;k}=0$), $g(u,\xi )=l:=0:\xi =\xi _{\perp }$. 
\begin{equation}  \label{Ch 8 2.40}
\pm \,l_{\xi _{\perp }}.(ul_{\xi _{\perp }})=g(_{rel}v,\xi _{\perp })\text{ .%
}
\end{equation}

In $(L_n,g)$-spaces the covariant derivative $\nabla _ug$ of the metric
tensor field $g$ along $u$ can be decomposed in its trace free part $%
^s\nabla _ug$ and its trace part $\frac 1n.Q_u.g$ as 
\[
\nabla _ug=\,^s\nabla _ug+\frac 1n.Q_u.g\text{ ,\thinspace \thinspace
\thinspace \thinspace \thinspace \thinspace \thinspace \thinspace \thinspace 
}\dim M=n\text{ ,} 
\]

\noindent where $\overline{g}[^s\nabla _ug]=0$,\thinspace \thinspace
\thinspace \thinspace \thinspace \thinspace \thinspace $Q_u=\overline{g}[%
\nabla _ug]=g^{kl}.g_{kl;j}.u^j=Q_j.u^j$, \thinspace \thinspace \thinspace
\thinspace $Q_j=g^{kl}.g_{kl;j}$.

\textit{Remark}. The covariant vector $\overline{Q}=\frac 1n.Q=\frac
1n.Q_j.dx^j=\frac 1n.Q_\alpha .e^\alpha $ is called 
\index{Weyl's vector field@Weyl's covector field} \textit{Weyl's covector
field}. The operator $^s\nabla _u=\,\nabla _u-\frac 1n.Q_u$ is called 
\textit{trace free covariant operator}.

If we use now the decomposition of $\nabla _ug$ in the expression for $\pm
\,2.l_\xi .(ul_\xi )$ we find the relation 
\begin{equation}  \label{Ch 8 2.41}
\begin{array}{c}
\pm \,2.l_\xi .(ul_\xi )=(^s\nabla _ug)(\xi ,\xi )+\frac 1n.Q_u.l_\xi
^2+2g(\nabla _u\xi ,\xi )= \\ 
=(^s\nabla _ug)(\xi _{\perp },\xi _{\perp })+ \\ 
+\frac le.[2.(^s\nabla _ug)(\xi _{\perp },u)+2.g(\nabla _u\xi ,u)+\frac
le.(^s\nabla _ug)(u,u)]+ \\ 
+\frac 1n.Q_u.(l_{\xi _{\perp }}^2+%
\frac{l^2}e)+2.g(_{rel}v,\xi _{\perp })\text{ ,}
\end{array}
\end{equation}

\noindent where $l_{\xi _{\perp }}^2=g(\xi _{\perp },\xi _{\perp })$, $%
l=g(\xi ,u)$.

For $l_\xi \neq 0:$%
\begin{equation}  \label{Ch 8 2.42}
ul_\xi =\pm \frac 1{2.l_\xi }(^s\nabla _ug)(\xi ,\xi )\pm \frac
1{2.n}.Q_u.l_\xi \pm \frac 1{l_\xi }.g(\nabla _u\xi ,\xi )\text{ .}
\end{equation}

In the case of a parallel transport ($\nabla _u\xi =0$) of $\xi $ along $u$
the change $ul_\xi $ of the length\thinspace \thinspace $l_\xi $ is 
\begin{equation}  \label{Ch 8 2.43}
ul_\xi =\pm \frac 1{2.l_\xi }(^s\nabla _ug)(\xi ,\xi )\pm \frac
1{2.n}.Q_u.l_\xi \text{ .\thinspace \thinspace \thinspace \thinspace
\thinspace \thinspace \thinspace \thinspace \thinspace \thinspace \thinspace
\thinspace \thinspace \thinspace \thinspace }
\end{equation}

\textit{Special case}: $\nabla _u\xi =0$ and $^s\nabla _ug=0$. 
\begin{equation}  \label{Ch 8 2.43a}
ul_\xi =\pm \frac 1{2.n}.Q_u.l_\xi \text{ .\thinspace \thinspace }
\end{equation}

\textit{Special case}: $g(u,\xi )=l:=0:\xi =\xi _{\perp }$. 
\[
\pm 2.l_{\xi _{\perp }}.(ul_{\xi _{\perp }})=(^s\nabla _ug)(\xi _{\perp
},\xi _{\perp })+\frac 1n.Q_u.l_{\xi _{\perp }}^2+2.g(_{rel}v,\xi _{\perp })%
\text{ .} 
\]
\begin{equation}  \label{Ch 8 2.44}
ul_{\xi _{\perp }}=\pm \frac 1{2.l_{\xi _{\perp }}}.(^s\nabla _ug)(\xi
_{\perp },\xi _{\perp })\pm \frac 1{2n}.Q_u.l_{\xi _{\perp }}\pm \frac
1{l_{\xi _{\perp }}}.g(_{rel}v,\xi _{\perp })\text{ ,\thinspace \thinspace
\thinspace \thinspace \thinspace \thinspace \thinspace \thinspace \thinspace
\thinspace }l_{\xi _{\perp }}\neq 0\text{ .}
\end{equation}

\textit{Special case}: Quasi-metric transports: $\nabla _ug:=2.g(u,\eta ).g$%
, \thinspace \thinspace \thinspace $u$, $\eta \in T(M)$. 
\begin{equation}  \label{Ch 8 2.45}
\pm 2.l_\xi .(ul_\xi )=2.g(u,\eta ).(l_{\xi _{\perp }}^2+\frac{l^2}%
e)+2.[\frac le.g(\nabla _u\xi ,u)+g(_{rel}v,\xi _{\perp })]\text{ .}
\end{equation}

\subsection{Change of the cosine between two contravariant vector fields and
the relative velocity}

The cosine between two contravariant vector fields $\xi $ and $\eta $ has
been defined as $g(\xi ,\eta )=l_\xi .l_\eta .\cos (\xi ,\eta )$. The rate
of change of the cosine along a contravariant vector field $u$ can be found
in the form 
\[
\begin{array}{c}
l_\xi .l_\eta .\{u[\cos (\xi ,\eta )]\}=(\nabla _ug)(\xi ,\eta )+g(\nabla
_u\xi ,\eta )+g(\xi ,\nabla _u\eta )- \\ 
-[l_\eta .(ul_\xi )+l_\xi .(ul_\eta )].\cos (\xi ,\eta )\text{ .}
\end{array}
\]

\textit{Special case}: $\nabla _u\xi =0$, $\nabla _u\eta =0$, $^s\nabla
_ug=0 $. 
\[
l_\xi .l_\eta .\{u[\cos (\xi ,\eta )]\}=\frac 1n.Q_u.g(\xi ,\eta )-[l_\eta
.(ul_\xi )+l_\xi .(ul_\eta )].\cos (\xi ,\eta )\text{ .} 
\]

Since $g(\xi ,\eta )=l_\xi .l_\eta .\cos (\xi ,\eta )$, it follows from the
last relation 
\[
l_\xi .l_\eta .\{u[\cos (\xi ,\eta )]\}=\{\frac 1n.Q_u.l_\xi .l_\eta
-[l_\eta .(ul_\xi )+l_\xi .(ul_\eta )]\}.\cos (\xi ,\eta )\text{ .} 
\]

Therefore, if $\cos (\xi ,\eta )=0$ between two parallel transported along $%
u $ vector fields $\xi $ and $\eta $, then the right angle between them
[determined by the condition $\cos (\xi ,\eta )=0$] does not change along
the contravariant vector field $u$. In the cases, when $\cos (\xi ,\eta
)\neq 0$, the rate of change of the cosine of the angle between two vector
fields $\xi $ and $\eta $ is linear to $\cos (\xi ,\eta )$.

By the use of the definitions and the relations: 
\begin{equation}  \label{Ch 8 2.46}
_{rel}v_\xi :=\overline{g}[h_u(\nabla _u\xi )]=\,_{rel}v\text{ ,\thinspace
\thinspace \thinspace \thinspace \thinspace \thinspace \thinspace \thinspace
\thinspace \thinspace \thinspace \thinspace \thinspace \thinspace \thinspace
\thinspace \thinspace }_{rel}v_\eta :=\overline{g}[h_u(\nabla _u\eta )]\text{
,}
\end{equation}
\begin{equation}  \label{Ch 8 2.47}
\begin{array}{c}
g(\nabla _u\xi ,\eta )=\frac 1e.g(u,\eta ).g(\nabla _u\xi ,u)+g(_{rel}v_\xi
,\eta ) \text{ ,} \\ 
g(\nabla _u\eta ,\xi )=\frac 1e.g(u,\xi ).g(\nabla _u\eta ,u)+g(_{rel}v_\eta
,\xi )\text{ ,}
\end{array}
\end{equation}
\begin{equation}  \label{Ch 8 2.48}
(\nabla _ug)(\xi ,\eta )=(^s\nabla _ug)(\xi ,\eta )+\frac 1n.Q_u.g(\xi ,\eta
)\text{ ,}
\end{equation}
\begin{equation}  \label{Ch 8 2.49}
\begin{array}{c}
(^s\nabla _ug)(\xi ,\eta )=(^s\nabla _ug)(\xi _{\perp },\eta _{\perp
})+\frac le.(^s\nabla _ug)(u,\eta _{\perp })+ \frac{\widetilde{l}}%
e.(^s\nabla _ug)(\xi _{\perp },u)+ \\ 
+\frac le.\frac{\widetilde{l}}e.(^s\nabla _ug)(u,u)\text{ ,\thinspace
\thinspace \thinspace \thinspace \thinspace \thinspace \thinspace \thinspace 
}\widetilde{l}=g(u,\eta )\text{ ,\thinspace \thinspace \thinspace \thinspace
\thinspace \thinspace \thinspace \thinspace }\eta _{\perp }=\overline{g}[%
h_u(\eta )]\text{ ,\thinspace \thinspace \thinspace \thinspace \thinspace }%
l=g(u,\xi )\text{ ,}
\end{array}
\end{equation}
\begin{equation}  \label{Ch 8 2.50}
\begin{array}{c}
(\nabla _ug)(\xi ,\eta )=(^s\nabla _ug)(\xi ,\eta )+\frac 1n.Q_u.g(\xi ,\eta
)= \\ 
=(^s\nabla _ug)(\xi _{\perp },\eta _{\perp })+\frac le.(^s\nabla _ug)(u,\eta
_{\perp })+ \frac{\widetilde{l}}e.(^s\nabla _ug)(\xi _{\perp },u)+ \\ 
+\frac le.\frac{\widetilde{l}}e.(^s\nabla _ug)(u,u)+\frac 1n.Q_u.[\frac{l.%
\widetilde{l}}e+g(\xi _{\perp },\eta _{\perp })]\text{ ,}
\end{array}
\end{equation}

\noindent the expression of $l_\xi .l_\eta .\{u[\cos (\xi ,\eta )]\}$
follows in the form 
\begin{equation}
\begin{array}{c}
l_\xi .l_\eta .\{u[\cos (\xi ,\eta )]\}=(^s\nabla _ug)(\xi _{\perp },\eta
_{\perp })+\frac le.[(^s\nabla _ug)(u,\eta _{\perp })+g(\nabla _u\eta ,u)]+
\\ 
+\frac{\widetilde{l}}e.[(^s\nabla _ug)(\xi _{\perp },u)+g(\nabla _u\xi ,u)]+%
\frac{l.\widetilde{l}}{e^2}.(^s\nabla _ug)(u,u)+ \\ 
+\,\frac 1n.Q_u.[\frac{l.\widetilde{l}}e+g(\xi _{\perp },\eta _{\perp
})]+g(_{rel}v_\xi ,\eta )+g(_{rel}v_\eta ,\xi )- \\ 
-[l_\eta .(ul_\xi )+l_\xi .(ul_\eta )].\cos (\xi ,\eta )\text{ .}
\end{array}
\label{Ch 8 2. 51}
\end{equation}

\textit{Special case}: $g(u,\xi )=l:=0$, $g(u,\eta )=\widetilde{l}:=0:\xi
=\xi _{\perp }$, $\eta =\eta _{\perp }$. 
\begin{equation}  \label{Ch 8 2.52}
\begin{array}{c}
l_{\xi _{\perp }}.l_{\eta _{\perp }}.\{u[\cos (\xi _{\perp },\eta _{\perp
})]\}=(^s\nabla _ug)(\xi _{\perp },\eta _{\perp })+\,\frac 1n.Q_u.l_{\xi
_{\perp }}.l_{\eta _{\perp }}.\cos (\xi _{\perp },\eta _{\perp })+ \\ 
+\,g(_{rel}v_{\xi _{\perp }},\eta _{\perp })+g(_{rel}v_{\eta _{\perp }},\xi
_{\perp })-[l_{\eta _{\perp }}.(ul_{\xi _{\perp }})+l_{\xi _{\perp
}}.(ul_{\eta _{\perp }})].\cos (\xi _{\perp },\eta _{\perp })\text{ ,}
\end{array}
\end{equation}

\noindent where $g(\xi _{\perp },\eta _{\perp })=l_{\xi _{\perp }}.l_{\eta
_{\perp }}.\cos (\xi _{\perp },\eta _{\perp })$.

\textit{Special case}: $^s\nabla _ug:=0:\nabla _ug=\frac 1n.Q_u.g$ (Weyl's
space with torsion). 
\begin{equation}  \label{Ch 9 2.62}
\pm 2.l_\xi .(ul_\xi )=2.\frac le.g(\nabla _u\xi ,u)+\frac 1n.Q_u.(l_{\xi
_{\perp }}^2+\frac{l^2}e)+2.g(_{rel}v,\xi _{\perp })\text{ ,}
\end{equation}
\begin{equation}  \label{Ch 9 2.63}
g(\nabla _u\xi ,u)=ul-\frac 1n.Q_u.l-g(\xi ,a)\text{ , \thinspace \thinspace
\thinspace \thinspace \thinspace \thinspace \thinspace \thinspace \thinspace
\thinspace \thinspace \thinspace \thinspace \thinspace \thinspace \thinspace
\thinspace \thinspace \thinspace \thinspace \thinspace \thinspace \thinspace
\thinspace \thinspace }a=\nabla _uu\text{ ,}
\end{equation}
\begin{equation}  \label{Ch 9 2.64}
\begin{array}{c}
l_\xi .l_\eta .\{u[\cos (\xi ,\eta )]\}=\frac le.g(\nabla _u\eta ,u)+ \frac{%
\widetilde{l}}e.g(\nabla _u\xi ,u)+ \\ 
+\,\frac 1n.Q_u.[ \frac{l.\widetilde{l}}e+g(\xi _{\perp },\eta _{\perp
})]+g(_{rel}v_\xi ,\eta )+g(_{rel}v_\eta ,\xi )- \\ 
-[l_\eta .(ul_\xi )+l_\xi .(ul_\eta )].\cos (\xi ,\eta )\text{ ,}
\end{array}
\end{equation}

\textit{Special case}: $^s\nabla _ug:=0:\nabla _ug=\frac 1n.Q_u.g$%
,\thinspace \thinspace \thinspace \thinspace \thinspace \thinspace
\thinspace $g(u,\xi )=l:=0$,\thinspace \thinspace \thinspace \thinspace
\thinspace \thinspace \thinspace \thinspace \thinspace \thinspace $g(u,\eta
)=\widetilde{l}:=0$ (Weyl's space with torsion and orthogonal to $u$ vector
fields $\xi _{\perp } $ and $\eta _{\perp }$). 
\begin{equation}  \label{Ch 9 2.66}
\pm 2.l_{\xi _{\perp }}.(ul_{\xi _{\perp }})=\frac 1n.Q_u.l_{\xi _{\perp
}}^2+2.g(_{rel}v,\xi _{\perp })\text{ ,}
\end{equation}
\begin{equation}  \label{Ch 9 2.67}
\begin{array}{c}
l_{\xi _{\perp }}.l_{\eta _{\perp }}.\{u[\cos (\xi _{\perp },\eta _{\perp
})]\}=\frac 1n.Q_u.g(\xi _{\perp },\eta _{\perp })+ \\ 
+\,\,g(_{rel}v_{\xi _{\perp }},\eta _{\perp })+g(_{rel}v_{\eta _{\perp
}},\xi _{\perp })-[l_{\eta _{\perp }}.(ul_{\xi _{\perp }})+l_{\xi _{\perp
}}.(ul_{\eta _{\perp }})].\cos (\xi _{\perp },\eta _{\perp }) \text{ }= \\ 
=g(_{rel}v_{\xi _{\perp }},\eta _{\perp })+g(_{rel}v_{\eta _{\perp }},\xi
_{\perp })- \\ 
-\,[l_{\eta _{\perp }}.(ul_{\xi _{\perp }})+l_{\xi _{\perp }}.(ul_{\eta
_{\perp }})+\frac 1n.Q_u.l_{\xi _{\perp }}.l_{\eta _{\perp }}].\cos (\xi
_{\perp },\eta _{\perp })
\end{array}
\end{equation}

\textit{Special case}: $U_n$-spaces: $\nabla _\xi g=0$ for $\forall \xi \in
T^{*}(M)$: 
\[
\begin{array}{c}
l_{\xi _{\perp }}.l_{\eta _{\perp }}.\{u[\cos (\xi _{\perp },\eta _{\perp
})]\}= \\ 
=g(_{rel}v_{\xi _{\perp }},\eta _{\perp })+g(_{rel}v_{\eta _{\perp }},\xi
_{\perp })-[l_{\eta _{\perp }}.(ul_{\xi _{\perp }})+l_{\xi _{\perp
}}.(ul_{\eta _{\perp }})].\cos (\xi _{\perp },\eta _{\perp }).
\end{array}
\]

\subsection{Rate of change of the length of a vector field $\xi _{\perp }$
connecting two particles (points) in the space-time as a $U_n$-space}

The distance between a particle (as a basic point) and an other particle (as
an observed point) lying in a neighborhood of the first one can be
determined by the use of the length of the vector field $\xi _{\perp }$. The
rate of change of the length $l_{\xi _{\perp }}$ of the vector field $\xi
_{\perp }$ (along the vector field $u$) in a $U_n$-space can be given in the
form 
\[
ul_{\xi _{\perp }}=\pm \,\frac 1{l_{\xi _{\perp }}}.g(\nabla _u\xi _{\perp
},\xi _{\perp })\text{ ,\thinspace \thinspace \thinspace \thinspace
\thinspace \thinspace \thinspace }l_{\xi _{\perp }}\neq 0\text{ .} 
\]

If $u$ is tangential vector field along a congruence of curves with
parameter $s$, i. e. if $u=\frac d{ds}$, then $ul_{\xi _{\perp }}$ will have
the form 
\[
\frac{dl_{\xi _{\perp }}}{ds}=\pm \,\frac 1{l_{\xi _{\perp }}}.g(\nabla
_u\xi _{\perp },\xi _{\perp })=\pm \,\frac 1{l_{\xi _{\perp }}}.g_{ij}.\xi
_{\perp }^i\,_{;k}.u^k.\xi _{\perp }^j\text{ .} 
\]

By the use of the decomposition of $\nabla _u\xi _{\perp }$, 
\[
\nabla _u\xi _{\perp }=\frac{\overline{l}}e.u+\,_{rel}v\,\,\,\,\,\,\,\text{,
\thinspace \thinspace \thinspace \thinspace \thinspace \thinspace }\overline{%
l}=g(\nabla _u\xi _{\perp },u)\text{ ,} 
\]

\noindent under the condition for the orthogonality between $u$ and $\xi
_{\perp }:g(u,\xi _{\perp })=l=0$, we obtain 
\[
\frac{dl_{\xi _{\perp }}}{ds}=\pm \,\frac 1{l_{\xi _{\perp }}}.g(_{rel}v,\xi
_{\perp })\text{ .} 
\]

Under the assumption $\pounds _{\xi _{\perp }}u=-\,\pounds _u\xi _{\perp }=0$%
, it follows that 
\begin{equation}  \label{4}
\frac{dl_{\xi _{\perp }}}{ds}=\pm \,\frac 1{l_{\xi _{\perp }}}.g(_{rel}v,\xi
_{\perp })=\pm \,\frac 1{l_{\xi _{\perp }}}.d(\xi _{\perp },\xi _{\perp })%
\text{ .}
\end{equation}

If we use the explicit form for $d$ and the fact that $\omega (\xi _{\perp
},\xi _{\perp })=0$, then we can find the expression for the rate of change
of $l_{\xi _{\perp }}$ in the form 
\begin{equation}  \label{5}
\frac{dl_{\xi _{\perp }}}{ds}=\pm \,\frac 1{l_{\xi _{\perp }}}.d(\xi _{\perp
},\xi _{\perp })=\pm \,\frac 1{l_{\xi _{\perp }}}.\sigma (\xi _{\perp },\xi
_{\perp })\pm \frac 1{n-1}.\theta .l_{\xi _{\perp }}\text{ .}
\end{equation}

\textit{Remark. }The sign $\pm $ depends on the sign of the metric $g$ (for $%
n=4 $, sign $g=\pm 2$).

Since the existence of the torsion could cause the condition $d=0$,
respectively $\sigma =0$, and $\theta =0$, the length $l_{\xi _{\perp }}$
would not change along $u$ and therefore, along the proper time $\tau =\frac
sc$, if we consider $\tau $ as the proper time of the basic particle (point).

\section{Relative acceleration and its kinematic characteristics induced by
the torsion}

\subsection{Relative acceleration in $(L_n,g)$-spaces}

The notion 
\index{relative acceleration@relative acceleration} \textit{relative
acceleration} vector field (relative acceleration) $_{rel}a$ in $(L_n,g)$%
-spaces can be defined (in analogous way as $_{rel}v$) (regardless of its
physical interpretation) as the orthogonal to a non-null contravariant
vector field $u$ [$g(u,u)=e\neq 0$] projection of the second covariant
derivative (along the same non-null vector field $u$) of (another)
contravariant vector field $\xi $, i.e. 
\begin{equation}  \label{Ch 9 3.1}
_{rel}a=%
\overline{g}(h_u(\nabla _u\nabla _u\xi ))=g^{ij}.h_{jk}.(\xi ^k\text{ }%
_{;l}.u^l)_{;m}.u^m.e_i\text{ .}
\end{equation}

$\nabla _u\nabla _u\xi $ $=(\xi ^i$ $_{;l}.u^l)_{;m}.u^m.e_i$ is the second
covariant derivative of a vector field $\xi $ along the vector field $u$. It
is an essential part of all types of deviation equations in $V_n$- and ($%
L_n,g$)-spaces \cite{Manoff-10} 
\index{Manoff S.@Manoff S.}, \cite{Iliev-8}%
\index{Iliev B.@Iliev B.}%
\index{Manoff S. (s. Iliev B.)@Manoff S. (s. Iliev B.)}.

If we take into account the expression for $\nabla _u\xi :\nabla _u\xi
=k[g(\xi )]-\pounds _\xi u$, and differentiate covariant along $u$, then we
obtain 
\[
\nabla _u\nabla _u\xi =\{\nabla _u[(k)g]\}(\xi )+(k)(g)(\nabla _u\xi
)-\nabla _u(\pounds _\xi u)%
\text{ .} 
\]

By means of the relations 
\[
\begin{array}{c}
k(g) \overline{g}=k\text{ , \thinspace \thinspace \thinspace \thinspace
\thinspace \thinspace \thinspace \thinspace \thinspace \thinspace \thinspace
\thinspace \thinspace \thinspace \thinspace \thinspace \thinspace \thinspace 
}\nabla _u[k(g)]=(\nabla _uk)(g)+k(\nabla _ug)\text{ ,} \\ 
\text{ }\{\nabla _u[k(g)]\}\overline{g}=\nabla _uk+k(\nabla _ug)\overline{g}%
\text{ ,}
\end{array}
\]

\noindent $\nabla _u\nabla _u\xi $ can be written in the form 
\begin{equation}
\nabla _u\nabla _u\xi =\frac le.H(u)+B(h_u)\xi -k(g)\pounds _\xi u-\nabla
_u(\pounds _\xi u)  \label{Ch 9 3.2}
\end{equation}

\noindent [compare with $\nabla _u\xi =\frac le.a+k(h_u)\xi -\pounds _\xi u$%
],

\noindent where 
\[
H=B(g)=(\nabla _uk)(g)+k(\nabla _ug)+k(g)k(g)\text{ ,} 
\]
\[
B=\nabla _uk+k(g)k+k(\nabla _ug)\overline{g}=\nabla _uk+k(g)k-k(g)(\nabla _u%
\overline{g})\text{ .} 
\]

The orthogonal to $u$ covariant projection of $\nabla _u\nabla _u\xi $ will
have therefore the form 
\begin{equation}  \label{Ch 9 3.3}
h_u(\nabla _u\nabla _u\xi )=h_u[\frac leH(u)-k(g)\pounds _\xi u-\nabla
_u\pounds _\xi u]+[h_u(B)h_u](\xi )\text{ .}
\end{equation}

In the special case, when $g(u,\xi )=l=0$ and $\pounds _\xi u=0$ , the above
expression has the simple form 
\begin{equation}  \label{Ch 9 3.4}
g(_{rel}a)=h_u(\nabla _u\nabla _u\xi )=[h_u(B)h_u](\xi )=A(\xi )\text{ ,}
\end{equation}

\noindent [compare with $h_u(\nabla _u\xi )=[h_u(k)h_u](\xi )=d(\xi )$].

The explicit form of $H(u)$ follows from the explicit form of $H$ and its
action on the vector field $u$%
\begin{equation}  \label{Ch 9 3.5}
H(u)=(\nabla _uk)[g(u)]+k(\nabla _ug)(u)+k(g)(a)=\nabla _u[k(g)(u)]=\nabla
_ua\text{ .}
\end{equation}

Now $h_u[\nabla _u\nabla _u\xi ]$ can be written in the form 
\begin{equation}  \label{Ch 9 3.6}
h_u(\nabla _u\nabla _u\xi )=h_u[\frac le.\nabla _ua-k(g)(\pounds _\xi
u)-\nabla _u(\pounds _\xi u)]+A(\xi )
\end{equation}

\noindent [compare $h_u(\nabla _u\xi )=h_u(\frac le.a-\pounds _\xi u)+d(\xi
) $].

The explicit form of $A=h_u(B)h_u$ can be found in an analogous way as the
explicit form for $d=h_u(k)h_u$ in the expression for $_{rel}v$.

\subsection{Deformation acceleration, shear acceleration, rotation
acceleration and expansion acceleration}

The covariant tensor $A$, named 
\index{deformation@deformation!deformation acceleration@deformation
acceleration} \textit{deformation acceleration} tensor can be represented as
a sum, containing three terms: a trace-free symmetric term, an antisymmetric
term and a trace term 
\begin{equation}  \label{Ch 9 3.7}
A=\,_sD+W+\frac 1{n-1}.U.h_u
\end{equation}

\noindent where 
\begin{equation}
D=h_u(_sB)h_u%
\text{ ,\thinspace \thinspace \thinspace \thinspace \thinspace \thinspace
\thinspace \thinspace \thinspace \thinspace }W=h_u(_aB)h_u\text{ ,\thinspace
\thinspace \thinspace \thinspace \thinspace \thinspace \thinspace \thinspace
\thinspace \thinspace \thinspace \thinspace }U=\overline{g}[_sA]=\overline{g%
}[D]  \label{Ch 9 3.8}
\end{equation}
\begin{equation}
_sB=\frac 12(B^{ij}+B^{ji}).e_i.e_j\text{ , \thinspace \thinspace \thinspace
\thinspace \thinspace \thinspace \thinspace \thinspace \thinspace \thinspace 
}_aB=\frac 12(B^{ij}-B^{ji}).e_i\wedge e_j\text{,}  \label{Ch 9 3.11}
\end{equation}
\begin{equation}
_sD=D-\frac 1{n-1}.\overline{g}[D].h_u=D-\frac 1{n-1}.U.h_u\text{ .}
\label{Ch 9 3.13}
\end{equation}

The tensor $_sD$ is the 
\index{shear@shear!shear acceleration@shear acceleration} \textit{shear
acceleration }tensor (shear acceleration), $W$ is the 
\index{rotation@rotation!rotation acceleration@rotation acceleration} 
\textit{rotation acceleration} tensor (rotation acceleration) and $U$ is the 
\index{expansion@expansion!expansion acceleration@expansion acceleration} 
\textit{expansion acceleration }invariant (expansion acceleration).
Furthermore, every one of these quantities can be divided into three parts:
torsion- and curvature-free acceleration, acceleration induced by the
torsion and acceleration induced by the curvature.

Let us now consider the representation of every acceleration quantity in its
essential parts connected with its physical interpretation.

The deformation acceleration tensor $A$ can be written in the following
forms 
\begin{equation}  \label{Ch 9 3.14}
A=\,_sD+W+\frac 1{n-1}.U.h_u=A_0+G=\,_FA_0-\,_TA_0+G%
\text{ ,}
\end{equation}
\begin{equation}  \label{Ch 9 3.15}
A=\,_sD_0+W_0+\frac 1{n-1}.U_0.h_u+\,_sM+N+\frac 1{n-1}.I.h_u\text{ ,}
\end{equation}
\begin{equation}  \label{Ch 9 3.16}
\begin{array}{c}
A=\,_{sF}D_0+\,_FW_0+\frac 1{n-1}._FU_0.h_u-(_{sT}D_0+\,_TW_0+\frac
1{n-1}._TU_{0.}h_u)+ \\ 
+\,_sM+N+\frac 1{n-1}.I.h_u\text{ ,}
\end{array}
\end{equation}

\noindent where $A_0$ is curvature-free deformation acceleration tensor, $G$
is the deformation acceleration tensor induced by the curvature, $_FA_0$ is
the torsion-free and curvature-free deformation acceleration tensor, $_TA_0$
is the deformation acceleration tensor induced by the torsion. Every of
these tensors could be represented in its three parts (the symmetric
trace-free part, the antisymmetric part and the trace part) determining the
corresponding shear acceleration, rotation acceleration and expansion
acceleration: 
\begin{equation}
A_0=\,_FA_0-\,_TA_0=\,_sD_0+W_0+\frac 1{n-1}.U_0.h_u\text{ ,}
\label{Ch 9 3.17}
\end{equation}
\begin{equation}
_FA_0=\,_{sF}D_0+\,_FW_0+\frac 1{n-1}._FU_0.h_u\text{ , \thinspace
\thinspace \thinspace \thinspace \thinspace \thinspace \thinspace }_FA_0(\xi
)=h_u(\nabla _{\xi _{\perp }}a)\text{ ,\thinspace \thinspace \thinspace }
\label{Ch 9 3.18}
\end{equation}
\begin{equation}
_TA_0=\,_{sT}D_0+\,_TW_0+\frac 1{n-1}._TU_0.h_u\text{ ,}  \label{Ch 9 3.20}
\end{equation}
\begin{equation}
G=\,_sM+N+\frac 1{n-1}.I.h_u=h_u(K)h_u\text{ ,}  \label{Ch 9 3.21}
\end{equation}
\begin{equation}
h_u([R(u,\xi )]u)=h_u(K)h_u(\xi )\text{ for }\forall \text{ }\xi \in T(M)%
\text{ , }  \label{Ch 9 3.22}
\end{equation}
\begin{equation}
\lbrack R(u,\xi )]u=\nabla _u\nabla _\xi u-\nabla _\xi \nabla _uu-\nabla
_{\pounds _u\xi }u\text{ ,}  \label{Ch 9 3.23}
\end{equation}
\begin{equation}
K=K^{kl}.e_k\otimes e_l\text{ , }K^{kl}=R^k\text{ }_{mnr}.g^{rl}.u^m.u^n%
\text{ ,}  \label{Ch 9 2.24}
\end{equation}

The components $R^k$ $_{mnr}$ are the components of the contravariant
Riemannian curvature tensor, 
\begin{equation}
K_a=K_a^{kl}.e_k\wedge e_l\text{, }K_a^{kl}=\frac 12(K^{kl}-K^{lk})\text{ , }%
K_s=K_s^{kl}.e_k.e_l\text{ , }K_s^{kl}=\frac 12(K^{kl}+K^{lk})\text{ ,}
\label{Ch 9 2.25}
\end{equation}
\begin{equation}
_sD=\,_sD_0+\,_sM\text{ , }W=W_0+N=\,_FW_0-\,_TW_0+N\text{ ,}
\label{Ch 9 2.26}
\end{equation}
\begin{equation}
U=U_0+I=\,_FU_0-\,_TU_0+I\text{ ,}  \label{Ch 9 2.27}
\end{equation}
\begin{equation}
_sM=M-\frac 1{n-1}.I.h_u\text{ ,\thinspace \thinspace \thinspace }%
M=h_u(K_s)h_u\text{ ,\thinspace \thinspace \thinspace }I=\overline{g}[M]=g^{%
\overline{i}\overline{j}}.M_{ij}\text{ ,}  \label{Ch 9 2.28}
\end{equation}
\begin{equation}
N=h_u(K_a)h_u\text{ ,}  \label{Ch 9 2.29}
\end{equation}
\begin{equation}
_sD_0=\,_{sF}D_0-\,_{sT}D_0=\,_FD_0-\frac 1{n-1}._FU_0.h_u-(_TD_0-\frac
1{n-1}._TU_0.h_u)\text{ ,}  \label{Ch 9 2.30}
\end{equation}
\begin{equation}
_sD_0=\,_{sF}D_0-\,_TD_0-\frac 1{n-1}(_FU_0-_TU_0)h_u\text{ ,}
\label{Ch 9 2.31}
\end{equation}
\begin{equation}
_sD_0=D_0-\frac 1{n-1}.U_0.h_u\text{ ,}  \label{Ch 9 2.32}
\end{equation}
\begin{equation}
_{sF}D_0=\,_FD_0-\frac 1{n-1}._FU_0.h_u\text{ , }_FD_0=h_u(b_s)h_u\text{ ,}
\label{Ch 9 2.33}
\end{equation}
\begin{equation}
b=b_s+b_a\text{ , \thinspace \thinspace \thinspace \thinspace \thinspace
\thinspace }b=b^{kl}.e_k\otimes e_l\text{ , \thinspace \thinspace \thinspace
\thinspace \thinspace \thinspace }b^{kl}=a^k\text{ }_{;n}.g^{nl}\text{ ,}
\label{Ch 9 2.34}
\end{equation}
\begin{equation}
a^k=u^k\text{ }_{;m}.u^m\text{ ,\thinspace \thinspace \thinspace \thinspace
\thinspace }b_s=b_s^{kl}.e_k.e_l\text{ , \thinspace \thinspace \thinspace
\thinspace \thinspace \thinspace \thinspace \thinspace }b_s^{kl}=\frac
12(b^{kl}+b^{lk})\text{ ,}  \label{Ch 9 2.35}
\end{equation}
\begin{equation}
b_a=b_a^{kl}.e_k\wedge e_l\text{ , \thinspace \thinspace \thinspace
\thinspace \thinspace \thinspace }b_a^{kl}=\frac 12(b^{kl}-b^{lk})\text{ ,}
\label{Ch 9 2.36}
\end{equation}
\begin{equation}
_FU_0=\overline{g}[_FD_0]=g[b]-\frac 1e.g(u,\nabla _ua)\text{ , }%
g[b]=g_{kl}.b^{kl}\text{ ,}  \label{Ch 9 2.37}
\end{equation}
\begin{equation}
_{sT}D_0=\,_TD_0-\frac 1{n-1}._TU_0.h_u=\,_{sF}D_0-\,_sD_0\text{ , }%
_TD_0=\,_FD_0-D_0\text{ ,}  \label{Ch 9 2.38}
\end{equation}
\begin{equation}
U_0=\overline{g}[D_0]=\,_FU_0-\,_TU_0\text{ ,\thinspace \thinspace
\thinspace \thinspace \thinspace \thinspace \thinspace \thinspace \thinspace
\thinspace }_TU_0=\overline{g}[_TD_0]\text{ ,}  \label{Ch 9 2.39}
\end{equation}
\begin{equation}
_FW_0=h_u(b_a)h_u\text{ , \thinspace \thinspace \thinspace \thinspace
\thinspace \thinspace \thinspace }_TW_0=\,_FW_0-W_0\text{ .}
\label{Ch 9 2.40}
\end{equation}

Under the conditions $\pounds _\xi u=0$ , $\xi =\xi _{\perp }=\overline{g}%
(h_u(\xi ))$ , ($l=0$), the expression for $h_u(\nabla _u\nabla _\upsilon
\xi )$ can be written in the forms 
\begin{equation}  \label{Ch 9 2.41}
g(_{rel}a)=h_u(\nabla _u\nabla _u\xi _{\perp })=A(\xi _{\perp })=A_0(\xi
_{\perp })+G(\xi _{\perp })\text{ ,}
\end{equation}
\begin{equation}  \label{Ch 9 2.42}
g(_{rel}a)=h_u(\nabla _u\nabla _u\xi _{\perp })=\,_FA_0(\xi _{\perp
})-\,_TA_0(\xi _{\perp })+G(\xi _{\perp })\text{ ,}
\end{equation}
\[
g(_{rel}a)=h_u(\nabla _u\nabla _u\xi _{\perp })=(_{sF}D_0+\,_FW_0+\frac
1{n-1}._FU_0.g)(\xi _{\perp })- 
\]
\begin{equation}  \label{Ch 9 2.43}
-(_{sT}D_0+\,_TW_0+\frac 1{n-1}._TU_0.g)(\xi _{\perp })+(_sM+N+\frac
1{n-1}.I.g)(\xi _{\perp })\text{ ,}
\end{equation}

\noindent which enable one to find a physical interpretation of the
quantities $_sD$,$W$,$U$ and of the quantities$_{sF}D_0$, $_FW_0$, $_FU_0$, $%
_{sT}D_0$, $_TW_0$, $_TU_0$, $_sM$, $N$, $I$ contained in their structure 
\cite{Manoff-9}.

After the above consideration the following proposition can be formulated:

\textit{Proposition 10}. The covariant vector field $g(_{rel}a)=h_u(\nabla
_u\nabla _u\xi )$ can be written in the form 
\[
g(_{rel}a)=h_u[\frac le.\nabla _ua-\nabla _{\pounds _\xi u}u-\nabla
_u(\pounds _\xi u)+T(\pounds _\xi u,u)]+A(\xi )\text{ ,} 
\]

\noindent with $A(\xi )=\,_sD(\xi )+W(\xi )+\frac 1{n\,-\,\,1}.U.h_u(\xi )$.

For the case of affine symmetric (Levi-Civita)%
\index{Levi-Civita connection@Levi-Civita connection} connection [$T(w,v)=0$
for $\forall $ $w,v\in T(M)$ , $T_{ij}\,^k=0$, $\Gamma _{ij}^k=\Gamma
_{ji}^k $] and Riemannian metric ($\nabla _vg=0$ for $\forall v\in T(M)$, $%
g_{ij;k}=0 $), kinematic characteristics are obtained in $V_n$-spaces
connected with the notion relative velocity \cite{Stephani} 
\index{Stephani H.@Stephani H.}, 
\index{Manoff S.@Manoff S.} and the relative acceleration \cite{Manoff-11} 
\cite{Manoff-12} 
\index{Manoff S.@Manoff S.}. For the case of affine non-symmetric connection
[$T(w,v)\neq 0$ for $\forall $ $w,v\in T(M)$ , $\Gamma _{jk}^i\neq \Gamma
_{kj}^i$] and Riemannian metric kinematic characteristics are obtained in $%
U_n$-spaces \cite{Manoff-11}%
\index{Manoff S.@Manoff S.}.

The shear, rotation and expansion accelerations induced by the torsion can
be expressed by means of the kinematic characteristics of the relative
velocity \cite{Manoff-9}:

(a) Shear acceleration tensor induced by the torsion%
\index{shear@shear!shear acceleration induced by the torsion@shear
acceleration induced by the torsion} $_{sT}D_0$%
\[
_{sT}D_0=\,_TD_0-\frac 1{n-1}._TU_0.h_u 
\]

\begin{equation}  \label{Ch 10 A.21}
\begin{array}{c}
_TD_0=\frac 12[_sP( 
\overline{g})\sigma +\sigma (\overline{g})_sP]+\frac 12[Q(\overline{g}%
)\omega +\omega (\overline{g})Q]+ \\ 
+\frac 1{n-1}(\theta _1.\sigma +\theta ._sP)+\frac 1{n-1}(\theta
_1^{.}+\frac 1{n-1}.\theta _1.\theta )h_u+\nabla _u(_sP)+ \\ 
+\frac 12[_sP( \overline{g})\omega -\omega (\overline{g})_sP]+\frac 12[Q(%
\overline{g})\sigma -\sigma (\overline{g})Q]+ \\ 
+\frac 1{2e}[h_u(a)\otimes (g(u))(m+q)h_u+h_u((g(u))(m+q))\otimes h_u(a)]+
\\ 
+\frac 1e[_sP(a)\otimes g(u)+g(u)\otimes _sP(a)]+ \\ 
+\frac 12[h_u(\nabla _u\overline{g})_sP+\,_sP(\nabla _u\overline{g}%
)h_u]+\frac 12[h_u(\nabla _ug)Q-Q(\nabla _ug)h_u]\text{ .}
\end{array}
\end{equation}

(b) Expansion acceleration induced by the torsion%
\index{expansion@expansion!expansion acceleration induced by the
torsion@expansion acceleration induced by the torsion} $_TU_0$ ($\theta
_1^{.}:=u\theta _1$) 
\begin{equation}  \label{Ch 10 A.23}
_TU_0=%
\overline{g}[_sP(\overline{g})\sigma ]+\overline{g}[Q(\overline{g})\omega %
]+\theta _1^{.}+\frac 1{n-1}.\theta _1.\theta +\frac 1e.g(u,T(a,u))\text{ .}
\end{equation}

(c) Rotation acceleration tensor induced by the torsion%
\index{rotation@rotation!rotation acceleration induced by the
torsion@rotation acceleration induced by the torsion} $_TW_0$%
\begin{equation}  \label{Ch 10 A.24}
\begin{array}{c}
_TW_0=\frac 12[_sP( 
\overline{g})\sigma -\sigma (\overline{g})_sP]+\frac 12[Q(\overline{g}%
)\omega -\omega (\overline{g})Q)]+ \\ 
+\frac 1{n-1}(\theta _1.\omega +\theta .Q)+\nabla _uQ+\frac 12[_sP( 
\overline{g})\omega +\omega (\overline{g})_sP]+ \\ 
+\frac 12[Q( \overline{g})\sigma +\sigma (\overline{g})Q]+ \\ 
+\frac 1{2e}[h_u(a)\otimes (g(u))(m+q)h_u-h_u((g(u))(m+q))\otimes h_u(a)]+
\\ 
+\frac 1e[Q(a)\otimes g(u)-g(u)\otimes Q(a)]+ \\ 
+\frac 12[h_u(\nabla _u\overline{g})_sP-\,_sP(\nabla _u\overline{g}%
)h_u]+\frac 12[h_u(\nabla _u\overline{g})Q+Q(\nabla _u\overline{g})h_u]\text{
.}
\end{array}
\end{equation}

The kinematic characteristics related to the notion of relative acceleration
can be used in finding out their influence on the rate of change of the
length of a contravariant vector field.

\subsection{Relative acceleration and change of the change of the length of
a contravariant vector field over a $(L_n,g)$-space}

The rate of change $ul_\xi $ of the length $l_\xi =\,\mid g(\xi ,\xi )\mid
^{\frac 12}$ of a contravariant vector field $\xi $ along a contravariant
vector field $u$ can be written in the form $\pm 2.l_\xi .(ul_\xi )=(\nabla
_ug)(\xi ,\xi )+2g(\nabla _u\xi ,\xi )$. After a further differentiation of
the last expression along the vector field $u$ we find the relation 
\begin{equation}  \label{Ch 9 2.45}
\begin{array}{c}
\pm (ul_\xi )^2\pm l_\xi .(u(ul_\xi ))=\frac 12.(\nabla _u\nabla _ug)(\xi
,\xi )+g(\nabla _u\nabla _u\xi ,\xi )+ \\ 
+2.(\nabla _ug)(\nabla _u\xi ,\xi )+g(\nabla _u\xi ,\nabla _u\xi )\text{ .}
\end{array}
\end{equation}

Our aim is to represent the terms of the right side of the last expression
by the use of the projective metrics of the vector field $u$ and the
decomposition of the metric tensor $g$ in its trace free and trace parts.
Since we already have the representation of this type for $ul_\xi $ (see the
above section), we will find the corresponding representation for $u(ul_\xi
)=uul_\xi $. After some computations, the following relations can be found:

(a) Representation of $(\nabla _u\nabla _ug)(\xi ,\xi )$. By the use of the
expressions:

\begin{equation}  \label{Ch 9 2.46}
^s\nabla _uQ_u=uQ_u-\frac 1n.Q_u^2\text{ ,}
\end{equation}
\begin{equation}  \label{Ch 9 2.47}
\nabla _u\nabla _ug=\,^s\nabla _u\,^s\nabla _ug+\frac
1n.[(uQ_u).g+2.Q_u.(^s\nabla _ug)]\text{ ,}
\end{equation}
\begin{equation}  \label{Ch 9 2.48}
(\nabla _u\nabla _ug)(\xi ,\xi )=(\nabla _u\nabla _ug)(\xi _{\perp },\xi
_{\perp })+\frac{l^2}{e^2}.(\nabla _u\nabla _ug)(u,u)+2.\frac le.(\nabla
_u\nabla _ug)(u,\xi _{\perp })\text{ ,}
\end{equation}
\begin{equation}  \label{Ch 9 2.49}
(\nabla _u\nabla _ug)(\xi _{\perp },\xi _{\perp })=(^s\nabla _u\,^s\nabla
_ug)(\xi _{\perp },\xi _{\perp })+\frac 2n.Q_u.(^s\nabla _ug)(\xi _{\perp
},\xi _{\perp })+\frac 1n.(uQ_u).l_{\xi _{\perp }}^2\text{ ,}
\end{equation}
\begin{equation}  \label{Ch 9 2.50}
\frac{l^2}{e^2}.(\nabla _u\nabla _ug)(u,u)=\frac{l^2}{e^2}.(^s\nabla
_u\,^s\nabla _ug)(u,u)+\frac 2n.\frac{l^2}{e^2}.Q_u.(^s\nabla
_ug)(u,u)+\frac 1n.\frac{l^2}e.(uQ_u)\text{ ,}
\end{equation}
\begin{equation}  \label{Ch 9 2.51}
(\nabla _u\,\nabla _ug)(u,\xi _{\perp })=(^s\nabla _u\,^s\nabla _ug)(u,\xi
_{\perp })+\frac 2n.Q_u.(^s\nabla _ug)(u,\xi _{\perp })\text{ ,}
\end{equation}

\noindent the expression for $(\nabla _u\nabla _ug)(\xi ,\xi )$ follows in
the form 
\begin{equation}
\begin{array}{c}
(\nabla _u\nabla _ug)(\xi ,\xi )=(^s\nabla _u\,^s\nabla _ug)(\xi _{\perp
},\xi _{\perp })+2.\frac le.(^s\nabla _u\,^s\nabla _ug)(u,\xi _{\perp })+ \\ 
+\frac{l^2}{e^2}.(^s\nabla _u\,^s\nabla _ug)(u,u)+\frac 2n.Q_u.[(^s\nabla
_ug)(\xi _{\perp },\xi _{\perp })+2.\frac le.(^s\nabla _ug)(u,\xi _{\perp
})]+ \\ 
+\frac 1n.(uQ_u).(l_{\xi _{\perp }}^2+\frac{l^2}e)+\frac 2n.\frac{l^2}{e^2}%
.Q_u.(^s\nabla _ug)(u,u)\text{ .}
\end{array}
\label{Ch 9 2.52}
\end{equation}

(b) Representation of $g(\nabla _u\nabla _u\xi ,\xi )$: 
\begin{equation}  \label{Ch 9 2.53}
g(\nabla _u\nabla _u\xi ,\xi )=g(_{rel}a,\xi _{\perp })+\frac le.g(u,\nabla
_u\nabla _u\xi )\text{ ,}
\end{equation}

(c) Representation of $(\nabla _ug)(\nabla _u\xi ,\xi )$. By the use of the
expressions 
\begin{equation}  \label{Ch 9 2.54}
\begin{array}{c}
(\nabla _ug)(\nabla _u\xi ,\xi )=\frac l{e^2}.(\nabla _ug)(u,u).g(\nabla
_u\xi ,u)+\frac 1e.(\nabla _ug)(u,\xi _{\perp }).g(\nabla _u\xi ,u)+ \\ 
+\,\frac le.(\nabla _ug)(_{rel}v,u)+(\nabla _ug)(_{rel}v,\xi _{\perp })\text{
,\thinspace \thinspace \thinspace \thinspace \thinspace \thinspace
\thinspace \thinspace \thinspace \thinspace \thinspace \thinspace \thinspace 
}_{rel}v=\overline{g}[h_u(\nabla _u\xi )]\text{ ,}
\end{array}
\end{equation}

\noindent and the representation of $\nabla _ug$ by the use of $^s\nabla _ug$
and $Q_u$, we obtain for $(\nabla _ug)(\nabla _u\xi ,\xi )$%
\begin{equation}
\begin{array}{c}
(\nabla _ug)(\nabla _u\xi ,\xi )=\frac 1e.(^s\nabla _ug)(u,\xi _{\perp
}).g(\nabla _u\xi ,u)+\frac l{e^2}.(^s\nabla _ug)(u,u).g(\nabla _u\xi ,u)+
\\ 
+\,(^s\nabla _ug)(_{rel}v,\xi _{\perp })+\frac le.(^s\nabla
_ug)(_{rel}v,u)+\frac 1n.Q_u.[g(_{rel}v,\xi _{\perp })+\frac le.g(\nabla
_u\xi ,u)]\text{ .}
\end{array}
\label{Ch 9 2.55}
\end{equation}

(d) Representation of $g(\nabla _u\xi ,\nabla _u\xi )$: 
\begin{equation}  \label{Ch 9 2.56}
g(\nabla _u\xi ,\nabla _u\xi )=g(_{rel}v,_{rel}v)+\frac 1e.[g(\nabla _u\xi
,u)]^2\text{ ,}
\end{equation}

\noindent where 
\begin{equation}
g(\nabla _u\xi ,\nabla _u\xi )=l_{\nabla _u\xi }^2\text{ ,\thinspace
\thinspace \thinspace \thinspace \thinspace \thinspace \thinspace \thinspace
\thinspace \thinspace }g(_{rel}v,_{rel}v)=l_{_{rel}v}^2\text{ , \thinspace
\thinspace \thinspace \thinspace \thinspace \thinspace \thinspace }%
e=g(u,u)=l_u^2\neq 0\text{ ,}  \label{Ch 9 2.57}
\end{equation}
\begin{equation}
g(\nabla _u\xi ,u)=ul-\frac 1n.Q_u.l-(^s\nabla _ug)(\xi ,u)-g(\xi ,a)\text{
,\thinspace \thinspace \thinspace \thinspace \thinspace \thinspace
\thinspace \thinspace \thinspace }a=\nabla _uu\text{ .}  \label{Ch 9 2.58}
\end{equation}

By the use of all main results from (a) - (d) we find the explicit form for $%
u(ul_\xi )$%
\begin{equation}  \label{Ch 9 2.59}
\begin{array}{c}
\pm \,\,l_\xi .(u(ul_\xi ))=\mp (ul_\xi )^2+\frac 12.[(^s\nabla _u\,^s\nabla
_ug)(\xi _{\perp },\xi _{\perp })+ \frac{l^2}{e^2}.(^s\nabla _u\,^s\nabla
_ug)(u,u)]+ \\ 
+\,\frac le.(^s\nabla _u\,^s\nabla _ug)(u,\xi _{\perp })+\frac
1n.Q_u.[(^s\nabla _ug)(\xi _{\perp },\xi _{\perp })+2.\frac le.(^s\nabla
_ug)(u,\xi _{\perp })+ \\ 
+\, \frac{l^2}{e^2}.(^s\nabla _ug)(u,u)+2.g(_{rel}v,\xi _{\perp })+2.\frac
le.g(\nabla _u\xi ,u)]+\frac 1{2n}.(uQ_u).(l_{\xi _{\perp }}^2+\frac{l^2}e)+
\\ 
+\frac 2e.(^s\nabla _ug)(u,\xi _{\perp }).g(\nabla _u\xi ,u)+2.\frac
l{e^2}.(^s\nabla _ug)(u,u).g(\nabla _u\xi ,u)+ \\ 
+2.(^s\nabla _ug)(_{rel}v,\xi _{\perp })+2.\frac le.(^s\nabla
_ug)(_{rel}v,u)+g(_{rel}v,_{rel}v)+\frac 1e.[g(\nabla _u\xi ,u)]^2+ \\ 
+\,g(_{rel}a,\xi _{\perp })+\frac le.g(\nabla _u\nabla _u\xi ,u)\text{ .}
\end{array}
\end{equation}

The last expression can help us to investigate the behavior of the length of
a contravariant vector filed $\xi $ when transported along a non-null
contravariant vector field $u$. There are several physical interpretation of
such type of a transport. If we interpret the vector field $u$ as a time
like vector field and as the 4-velocity of an observer in a $(L_n,g)$-space
considered as a model of the space-time, then $u(ul_\xi )=(d^2l_\xi )/(ds^2)$
represents the acceleration acting on the length $l_\xi $ of the vector $\xi 
$ along the observers trajectory $x^i(s)$. The vector field $u$ can also be
considered as a space like vector field with different from the above
physical interpretation.

\textit{Special case}: $g(u,\xi )=l:=0:\xi =\xi _{\perp }$. 
\begin{equation}  \label{Ch 9 2.60}
\begin{array}{c}
\pm \,\,l_{\xi _{\perp }}.(u(ul_{\xi _{\perp }}))=\mp (ul_{\xi _{\perp
}})^2+\frac 12.(^s\nabla _u\,^s\nabla _ug)(\xi _{\perp },\xi _{\perp })+ \\ 
+\,\frac 1n.Q_u.[(^s\nabla _ug)(\xi _{\perp },\xi _{\perp })+2.g(_{rel}v,\xi
_{\perp })]+\frac 1{2n}.(uQ_u).l_{\xi _{\perp }}^2+ \\ 
+\frac 2e.(^s\nabla _ug)(u,\xi _{\perp }).g(\nabla _u\xi _{\perp
},u)+2.(^s\nabla _ug)(_{rel}v,\xi _{\perp })+ \\ 
+g(_{rel}v,_{rel}v)+\frac 1e.[g(\nabla _u\xi _{\perp },u)]^2+\,g(_{rel}a,\xi
_{\perp })\text{ .}
\end{array}
\end{equation}

\textit{Special case}: $g(u,\xi )=l:=0:\xi =\xi _{\perp }$, $\nabla _\xi g=0$
for $\forall \xi \in T(M)$ [$g_{ij;k}:=0$] ($V_n$- or $U_n$-spaces). 
\begin{equation}  \label{Ch 9 2.61}
\pm \,\,l_{\xi _{\perp }}.(u(ul_{\xi _{\perp }}))=\mp (ul_{\xi _{\perp
}})^2+g(_{rel}v,_{rel}v)+\frac 1e.[g(\nabla _u\xi _{\perp
},u)]^2+\,g(_{rel}a,\xi _{\perp })\text{ .}
\end{equation}

\textit{Special case}: $^s\nabla _ug:=0:\nabla _ug=\frac 1n.Q_u.g$ (Weyl's
space with torsion). 
\begin{equation}  \label{Ch 9 2.65}
\begin{array}{c}
\pm l_\xi .(u(ul_\xi ))=\mp (ul_\xi )^2+\frac 2n.Q_u.[g(_{rel}v,\xi _{\perp
})+\frac le.g(\nabla _u\xi ,u)]+ \\ 
+\,\frac 1{2n}.(uQ_u).(l_{\xi _{\perp }}^2+ \frac{l^2}e)+g(_{rel}v,_{rel}v)+%
\frac 1e.[g(\nabla _u\xi ,u)]^2+g(_{rel}a,\xi _{\perp })+ \\ 
+\,\frac le.g(\nabla _u\nabla _u\xi ,u)\text{ .}
\end{array}
\end{equation}

\textit{Special case}: $^s\nabla _ug:=0:\nabla _ug=\frac 1n.Q_u.g$%
,\thinspace \thinspace \thinspace \thinspace \thinspace \thinspace
\thinspace $g(u,\xi )=l:=0$,\thinspace \thinspace \thinspace \thinspace
\thinspace \thinspace \thinspace \thinspace \thinspace \thinspace $g(u,\eta
)=\widetilde{l}:=0$ (Weyl's space with torsion and orthogonal to $u$ vector
fields $\xi _{\perp } $ and $\eta _{\perp }$). 
\begin{equation}  \label{Ch 9 2.68}
\begin{array}{c}
\pm l_{\xi _{\perp }}.(u(ul_{\xi _{\perp }}))=\mp (ul_{\xi _{\perp
}})^2+\frac 2n.Q_u.g(_{rel}v,\xi _{\perp })+ \\ 
+\,\frac 1{2n}.(uQ_u).l_{\xi _{\perp }}^2+g(_{rel}v,_{rel}v)+\frac
1e.[g(\nabla _u\xi _{\perp },u)]^2+g(_{rel}a,\xi _{\perp })\text{ .}
\end{array}
\end{equation}

The representations of $ul_\xi $, $u(ul_\xi )$ and $u[\cos (\xi ,\eta )]$
can be useful tools for the considerations of the motion of physical systems
with given dimensions in $(L_n,g)$-spaces. The induced by the torsion
kinematic characteristics of the relative velocity $_{rel}v$ and the
relative acceleration $_{rel}a$ have an equal position to the other
kinematic characteristics induced by the curvature or by external forces.

The relative acceleration between two (neighbor) points (particles) in an $%
U_n$-space $(n=4)$ is also related to quantities induced by non-autoparallel
motions (under external forces), by the curvature and by the torsion.

\subsection{Rate of change of the rate of change of the length of the vector
field $\xi _{\perp }$ over an $U_n$-space}

From the relation in $U_n$-spaces 
\[
ul_{\xi _{\perp }}=\pm \frac 1{l_{\xi _{\perp }}}.g(\nabla _u\xi _{\perp
},\xi _{\perp })\text{ ,\thinspace \thinspace \thinspace \thinspace
\thinspace \thinspace \thinspace \thinspace }l_{\xi _{\perp }}\neq 0\text{ ,}
\]

\noindent after covariant differentiation along the vector field $u$, we
obtain for $U_n$-spaces the relation 
\[
u(ul_{\xi _{\perp }})=\mp \frac 1{l_{\xi _{\perp }}^2}.(ul_{\xi _{\perp
}}).g(\nabla _u\xi _{\perp },\xi _{\perp })\pm \frac 1{l_{\xi _{\perp
}}}.g(\nabla _u\nabla _u\xi _{\perp },\xi _{\perp })\pm \frac 1{l_{\xi
_{\perp }}}.g(\nabla _u\xi _{\perp },\nabla _u\xi _{\perp })\text{ ,} 
\]

\noindent where 
\[
\mp \frac 1{l_{\xi _{\perp }}}.g(\nabla _u\xi _{\perp },\xi _{\perp })=\mp
ul_{\xi _{\perp }}\text{ ,\thinspace \thinspace \thinspace \thinspace }\mp
\frac 1{l_{\xi _{\perp }}^2}.(ul_{\xi _{\perp }}).g(\nabla _u\xi _{\perp
},\xi _{\perp })=\mp \frac 1{l_{\xi _{\perp }}^3}.[g(_{rel}v,\xi _{\perp
})]^2\text{ .} 
\]

By the use of the decomposition of $\nabla _u\nabla _u\xi _{\perp }$, 
\[
\nabla _u\nabla _u\xi _{\perp }=\frac{\widehat{l}}e.u+\,_{rel}a\text{
,\thinspace \thinspace \thinspace \thinspace \thinspace \thinspace }\widehat{%
l}=g(\nabla _u\nabla _u\xi _{\perp },u)\text{ ,} 
\]

\noindent under the condition $g(u,\xi _{\perp })=l=0$, we obtain 
\begin{equation}
u(ul_{\xi _{\perp }})=\mp \frac 1{l_{\xi _{\perp }}^3}.[g(_{rel}v,\xi
_{\perp })]^2\pm \frac 1{l_{\xi _{\perp }}}.g(_{rel}a,\xi _{\perp })\pm
\frac 1{l_{\xi _{\perp }}}.\frac{\overline{l}\,^2}e\pm \frac 1{l_{\xi
_{\perp }}}g(_{rel}v,_{rel}v)\text{ ,}  \label{28}
\end{equation}

\noindent where 
\[
\text{\thinspace \thinspace }\overline{l}=g(\nabla _u\xi _{\perp },u)=\nabla
_u[g(\xi _{\perp },u)]-(\nabla _ug)(\xi _{\perp },u)-g(\xi _{\perp },\nabla
_uu)\text{ .} 
\]

From the last relation, it follows for $U_n$-spaces $(\nabla _ug=0)$ and for 
$l=g(u,\xi _{\perp })=0$ that 
\[
\text{\thinspace \thinspace }\overline{l}=g(\nabla _u\xi _{\perp },u)=-g(\xi
_{\perp },\nabla _uu)=-g(\xi _{\perp },a)\text{ ,\thinspace \thinspace
\thinspace \thinspace \thinspace \thinspace }\nabla _uu=a\text{ .} 
\]

For $(L_n,g)$-spaces and $l=0$, it follows that 
\[
\text{\thinspace }\overline{l}=g(\nabla _u\xi _{\perp },u)=-(\nabla _ug)(\xi
_{\perp },u)-g(\xi _{\perp },\nabla _uu)\text{ .} 
\]

Then we obtain for $U_n$-spaces 
\begin{equation}  \label{32}
u(ul_{\xi _{\perp }})=\pm \,\frac 1{l_{\xi _{\perp }}}.\{g(_{rel}a,\xi
_{\perp })-\frac 1{l_{\xi _{\perp }}^2}[g(_{rel}v,\xi _{\perp })]^2+\frac
1e.[g(\xi _{\perp },a)]^2+g(_{rel}v,_{rel}v)\}\text{ . }
\end{equation}

Therefore, for $u=\frac d{ds}$ the relation follows 
\begin{equation}  \label{33}
\frac{d^2l_{\xi _{\perp }}}{ds^2}=\pm \,\frac 1{l_{\xi _{\perp
}}}.\{g(_{rel}a,\xi _{\perp })+g(_{rel}v,_{rel}v)+\frac 1e.[g(a,\xi _{\perp
})]^2-\frac 1{l_{\xi _{\perp }}^2}[g(_{rel}v,\xi _{\perp })]^2\}\text{ . }
\end{equation}

If $\pounds _u\xi _{\perp }=0$, then $_{rel}a=\overline{g}[A(\xi _{\perp })]$%
, $_{rel}v=\overline{g}[d(\xi _{\perp })]$. It is obvious that the existing
kinematic terms induced by the torsion in $A$ and in $d$ could compensate
the action of the kinematic terms induced by the curvature or by external
forces in such a way that no change of rate of the rate of change of $l_{\xi
_{\perp }}$ could be registered. If $_{rel}a=0$, $_{rel}v=0$ and $g(a,\xi
_{\perp })=0$, then $(d^2l_{\xi _{\perp }}/ds^2)=0$.

\section{Conclusion}

The recent development of the mathematical models of the space-time shows
the possible use of spaces with affine connections and metrics. Such type of
spaces have the torsion as an intrinsic characteristic. Even in (pseudo)
Riemannian spaces with torsion ($U_n$-spaces) the tensor of the torsion and
the kinematic characteristics related to it could influence under certain
conditions the effects in the space-time caused by the curvature tensor and
by external forces. Especially, the torsion could lead up to possible new
gravitational experiments. In such cases a consideration of the influence of
the torsion as a characteristic of the space-time appears as a necessary
step to a better understanding of the properties of the space-time related
to the gravitational interaction between physical systems.

\begin{center}
\textbf{Acknowledgments}
\end{center}

This work is supported in part by the National Science Foundation of
Bulgaria under Grant No. 642.

\small%


\begin{thebibliography}{99}
\bibitem{Hehl}  F. W. Hehl, J. D. McCrea, E. W. Mielke and Y. Ne'eman, 
\textit{Metric-affine gauge theory of gravity: field equations, Noether
identities, world spinors, and breaking of dilation invariance}. \textit{%
Physics Reports} \textbf{258} 1-2 1-171 (1995)

\bibitem{Iliev-1}  B. Z. Iliev B. Z., \textit{Special bases for derivations
of tensor algebras. I. Cases in a neighbourhood and at a point.} Comm. JINR
Dubna \textbf{E5-92-507 }(1992) 1-17

\bibitem{Iliev-2}  B. Z. Iliev\textbf{, }\textit{Special bases for
derivations of tensor algebras. II. Case along paths.}\textbf{\ }Comm. JINR
Dubna \textbf{E5-92-508} (1992) 1-16

\bibitem{Iliev-3}  B. Z. Iliev\textbf{, }\textit{Special bases for
derivations of tensor algebras. III. Case along smooth maps with separable
points of selfintersection}\textbf{\textit{.} }Comm. JINR Dubna \textbf{%
E5-92-543} (1992) 1-15

\bibitem{Iliev-4}  B. Z. Iliev, \textit{Normal frames and the validity of
the equivalence principle: I. Cases in a neighbourhood and at a point}. J.
Phys. A: Math. Gen. \textbf{29} (1996) 6895-6901

\bibitem{Iliev-5}  B. Z. Iliev, \textit{Normal frames and the validity of
the equivalence principle: II. The case along paths}. J. Phys. A: Math. Gen. 
\textbf{30} (1997) 4327-4336

\bibitem{Iliev-6}  B. Z. Iliev, \textit{Normal frames and the validity of
the equivalence principle: III. The case along smooth maps with separable
points of self-interaction}. J. Phys. A: Math. Gen. \textbf{31} (1998)
1287-1296

\bibitem{Iliev-7}  B. Z. Iliev, \textit{Is the principle of equivalence a
principle?} Journal of Geometry and Physics \textbf{24} (1998) 209-222

\bibitem{Hartley}  Hartley D., \textit{Normal frames for non-Riemannian
connections}. Class. and Quantum Grav. \textbf{12} (1995) L103-L105

\bibitem{Manoff-0}  S. Manoff, \textit{Spaces with contravariant and
covariant affine connections and metrics.} Physics of elementary particles
and atomic nucleus (Physics of Particles and Nuclei) [Russian Edition: 
\textbf{30} (1999) 5, 1211-1269], [English Edition: \textbf{30} (1999) 5,
527-549]

\bibitem{Manoff-01}  S. Manoff, \textit{Lagrangian theory of tensor fields
over spaces with contravariant and covariant affine connections and metrics
and its applications to Einstein's theory of gravitation in }$\overline{V}_4$%
-\textit{spaces.} Acta Appl. Math. \textbf{55} (1999) 1, 51-125

\bibitem{Manoff-1}  S. Manoff, \textit{Geodesic and autoparallel equations
over differentiable manifolds}. Intern. J. Mod. Phys. \textbf{A 11} (1996)
21, 3849-3874

\bibitem{Manoff-2}  S. Manoff, \textit{Auto-parallel equation as
Euler-Lagrange's equation over spaces with affine connections and metrics.}
Gen. Rel. and Grav. (2000) (to appear)

\bibitem{Manoff-3}  S. Manoff, \textit{Frames of reference in spaces with
affine connections and metrics}. E-print (1999) gr-qc/99 08 061

\bibitem{Manoff-4}  S. Manoff, \textit{Fermi-Walker transports over spaces
with affine connections and metrics}. JINR Rapid Communications \textbf{1
[81]} (1997) 5-12

\bibitem{Manoff-5}  S. Manoff, \textit{Fermi derivative and Fermi-Walker
transports over }$(L_n,g)$\textit{-spaces}. Class. Quantum Grav. \textbf{15}
(1998) 2, 465-477

\bibitem{Manoff-6}  S. Manoff, \textit{Fermi derivative and Fermi-Walker
transports over }$(\overline{L}_n,g)$\textit{-spaces}. Intern. J. Mod. Phys. 
\textbf{A 13} (1998) 25, 4289-4308

\bibitem{Manoff-7}  S. Manoff, \textit{Conformal derivative and conformal
transports over }$(L_n,g)$-spaces. E-print (1999) gr-qc/99 07 095

\bibitem{Manoff-8}  S. Manoff, \textit{Conformal derivative and conformal
transports over }$(\overline{L}_n,g)$-spaces. Intern. J. Mod. Phys. \textbf{%
A 15} (2000) (to appear)

\bibitem{Misner}  Misner Ch. W., Thorne K. S., Wheeler J. A., \textit{%
Gravitation. }(W. H. Freeman and Company, San Francisco, 1973), Ch. 37.
Russian translation: Vol 1., Vol. 2., Vol. 3. (Mir, Moscow, 1977)

\bibitem{Stephani}  Stephani H 1977 \textit{Allgemeine
Relativit\"{a}tstheorie}. (VEB Deutscher Verlag d. Wiss., Berlin) S 54-55.

\bibitem{Manoff-9}  S. Manoff, \textit{Kinematics of vector fields}. In 
\textit{Complex Structures and Vector Fields}. eds. Dimiev St., Sekigawa K.
(World Sci. Publ., Singapore, 1995), pp. 61-113

\bibitem{Grishchuk}  Grishchuk L P and Polnarev A G 1980 In \textit{General
Relativity and Gravitation} ed Held A Vol. 2. (Plenum Press, New York) pp
393-434.

\bibitem{Laemmerzahl}  Cl. L\"{a}mmerzahl, \textit{Constraints on space-time
torsion from Hyghes-Drever experiments}. Phys. Lett.A 228 223-231 (1997)

\bibitem{Yano}  Yano\ K 1957 \textit{The Theory of Lie Derivatives and its
Applications} (North Holland Publ. Co., Amsterdam)

\bibitem{Kramer}  Kramer D, Stephani H, MacCallum M, and Herlt E 1980 
\textit{Exact Solutions of Einstein's Field Equations}. (VEB Deutscher
Verlag d. Wiss., Berlin)

\bibitem{Synge}  Synge J L 1960 \textit{Relativity: the general theory}.
(North Holland Publ. Co., Amsterdam)

\bibitem{Ehlers}  Ehlers J., \textit{Beitr\"{a}ge zur relativistischen
Mechanik kontinuierlicher Medien}. Abhandlungen d. Mainzer Akademie d.
Wissenschaften, Math.-Naturwiss. Kl. Nr. \textbf{11 }(1961)

\bibitem{Manoff-10}  S. Manoff, \textit{Lie derivatives and deviation
equations in Riemannian spaces}.\textbf{\ }Gen. Rel. and Grav. \textbf{11 }%
(1979) 189-204

\bibitem{Iliev-8}  Iliev B., Manoff S., \textit{Deviation equations in
spaces with affine connection}. Comm. JINR Dubna \textbf{P2-83-897} (1983)
1-16

\bibitem{Manoff-11}  S. Manoff, \textit{Relative acceleration, shear,
rotation and expansion acceleration in (pseudo) Riemannian spaces with and
without torsion}. In \textit{Gravitational Waves.} JINR Dubna \textbf{%
P2-85-667} (1985) 157-168

\bibitem{Manoff-12}  S. Manoff, \textit{Deviation equations of Synge and
Schild in spaces with affine connection and metric, and equations for
gravitational waves detectors}. Comm. JINR Dubna \textbf{E2-92-19} (1992)
1-12
\end{thebibliography}
\end{document}